\documentclass[11pt]{article}
\usepackage[dvips]{graphicx}

\usepackage{color}
\usepackage[x11names,rgb]{xcolor}

\usepackage{amsfonts}
\usepackage{amsmath}
\usepackage{amssymb}
\usepackage{amsthm}
\usepackage{enumerate}
\usepackage{float}
\usepackage{caption}
\usepackage{subcaption}
\usepackage{siunitx}
\usepackage{booktabs}
\usepackage{eurosym}
\usepackage{caption}
\usepackage{subcaption}
\usepackage[pdftex]{hyperref}
\usepackage[round]{natbib}
\usepackage{lineno}
\usepackage{multirow}
\usepackage{arydshln}
\usepackage{lscape}
\usepackage{longtable}
\usepackage{authblk}
\usepackage{caption}
\usepackage{setspace}
\usepackage[para,online,flushleft]{threeparttable}
\usepackage{fancybox}

\theoremstyle{remark}

 {\end{color}}

\begin{document}
\title{From social interactions to private environmental behaviours: The case of consumer food waste}

\author[a]{Simone Piras}
\author[b]{Simone Righi}
\author[c]{Marco Setti}
\author[a]{Nazli Koseoglu}
\author[d]{Matthew J. Grainger}
\author[e]{Gavin B. Stewart}
\author[c]{Matteo Vittuari}

\affil[a]{Social, Economic and Geographical Sciences, The James Hutton Institute, Aberdeen AB15 8QH, Scotland, UK. Corresponding author. Email: \url{simone.piras@hutton.ac.uk}, Tel: +44 (0)1224 395 399}
\affil[b]{Department of Economics, Ca' Foscari University of Venice, Fondamenta S. Giobbe 873, 30121, Venezia, Italy.}
\affil[c]{Department of Agricultural and Food Sciences, Alma Mater Studiorum – University of Bolo-gna, Viale G. Fanin 50, 40137 Bologna, Italy.}
\affil[d]{Department of Terrestrial Ecology, Norwegian Institute for Nature Research, NO-7485
Trondheim, Norway.}
\affil[e]{School of Agriculture, Food and Rural Affairs, Agriculture Building, Newcastle University,
Newcastle upon Tyne NE1 7RU, UK.}

\date{}

\maketitle

\begin{abstract}
\noindent Consumer food waste, like many environmental behaviours, takes place in private, and is not directly subject to social monitoring. Nevertheless, social interactions can affect private opinions and behaviours. This paper builds an agent-based model of interactions between consumers heterogeneous in their sociability, initial opinions and behaviours related to food waste and willingness to consider different opinions, in order to assess how social interactions can affect private behaviours. Compared to existing models of opinion dynamics, we innovate by including a range of ``cognitive dissonance'' between stated opinions and actual behaviours that consumers are willing to accept before changing one of the two. We calibrate the model using questionnaire data on household food waste in Italy. We find that a limited degree of mixing between different socio-demographic groups, namely adult and young consumers, is enough to trigger change, but a certain openness of mind is required from more wasteful individuals. Equally, a small group of environmentally committed consumers can attract a sizeable share of the population towards low-waste behaviours if they show a certain variability of opinions and are willing to compromise with individuals in their close neighbourhood in terms of opinions. These findings can help design effective interventions to promote pro-environmental behaviours, taking advantage of the beneficial network effects while anticipating negative externalities.

\vspace{0.5cm}
\noindent \textbf{Keywords:} Food waste; social interactions; consumer opinion; consumer behaviour; cognitive dissonance; agent-based model.
\end{abstract}

\graphicspath{{figs/}}

\clearpage

This is the Accepted version of a paper with the same title that was published on October 10, 2021 in Resources, Conservation \& Recycling. The published version (and the supplementary material) is available at the address  \url{https://doi.org/10.1016/j.resconrec.2021.105952} and \url{https://www.sciencedirect.com/science/article/pii/S0921344921005619}.

\section{Introduction}
Food waste is a societal challenge drawing a burgeoning amount of attention in the agendas of public and private actors. In developed countries, consumers are responsible for the largest share of waste along value chains (\citealt{gustavsson2011global, gunders2012wasted, stenmarck2011initiatives}), equalling 53\% of the total food waste in the EU.\footnote{Retrieved from: \url{https://www.europarl.europa.eu/news/en/headlines/society/20170505STO73528/food-waste-the-problem-in-the-eu-in-numbers-infographic} [Accessed 11.02.2021].} This implies the waste of valuable resources like soil, energy and water, but also the unnecessary emission of huge amounts of CO\textsubscript{2} contributing to climate change. There has thus been an increasing focus on interventions to reduce food waste at the consumption stage. Most of the current literature on this matter focuses on the identification of socio-demographic and situational drivers (\citealt{wenlock1980household, sonesson2005home, wassermann2005edibles, barr2007factors, Wrap2014a, parizeau2015household, SettietAl2016, canali2016food}), and their relative influence (\citealt{grainger2018model}). A social component has also been identified (\citealt{evans2011blaming}), suggesting that food waste is not simply an individual problem. However, there has been limited attention to the interaction between idiosyncratic and social aspects of consumers' food waste behaviour, including though simulation models. An attempt to study the process of consumers' food waste generation through extensive agent-based simulations has been made by \cite{ravandi2019impact}. However, their agent-based model (ABM) focuses on logistic issues in a very specific setting (food-service operations), with limited role for social influence.

To fill this gap, this paper merges evidence from the food waste literature on the one side, and social simulation on the other side, to develop an ABM that explores how individual food waste behaviours are influenced by peers through exchanges of opinions. In particular, we study the impact on food waste of interactions between individuals belonging to different socio-demographic groups. First, we focus on the role of mixing between adult and young consumers; second, we explore the convincing power of a small group of individuals committed to food waste reduction. We pay particular attention to the mismatch between publicly stated opinions and private behaviours. Unlike most ABMs, which are calibrated qualitatively using general stylized facts, our model is partially calibrated with real-world data.

Our simulations show that peers can influence each other's behaviour even when this behaviour is highly heterogeneous and private. Networks and similarity emerge as moderators between social influence and behavioural change. Rather than overall convergence, we observe the emerging of different clusters, as behaviours tend to align with individuals from the same group. The case study of adult and young consumers, with the former wasting little and the latter wasting much, suggests that if individuals are sufficiently open-minded, small degrees of mixing between groups are enough to reduce the share of those with ``extreme'' behaviours. The case study of populations with a small group of committed individuals shows that, under certain conditions, these can convince a significant share of the overall population. Being the first model to assess the impact of social influence on food waste generation though extensive simulations, there is a lack of research for comparison. Nevertheless, our results suggest that policies addressing food waste should promote interaction opportunities among different socio-demographic groups to achieve larger impact.

The paper is structured as follows. First, we review the literature on food behaviours and social influence. Second, we describe our theoretical model and characterize our case studies starting from the literature and available data. Third, we present and discuss the results of extensive simulations. Finally, we draw our conclusions and illustrate their implications for research and policy design.

\section{Literature review}
Recent advancements in consumer studies have departed from the assumption that consumers make their decisions in isolation, based only on individual preferences and attitudes. The economics, psychology and sociology literature has been focusing on the impact of social embeddedness (i.e., of constraints due to the social context where economic activities take place) (\citealt{ jackson2005economics, jackson2010social,Jackson2016}), and peers' opinions (\citealt{galeotti2010talking, simpson2012relat, wood2012influence,lamberton2013destination,CarroniPinRighi}). When analysing the effects of social influence on behavioural change, social sciences identify two main moderators: existing networks (\citealt{abrahamse2013social}), and perceived similarity with others (social comparison theory \citealt{festinger1954theory}). As for existing networks, \cite{abelson1968theories} (quoted in \citealt{jackson2005economics}) show that individuals prefer to adopt behaviours coherent with their own and their peers' motivations. As for perceived similarity, a pattern coherent with the social comparison theory is observed when the reference group and the target group show similar characteristics (\citealt{goldstein2008room}), although this finding does not emerge in other parallel studies (\citealt{schultz2008using}). Moreover, when individuals identify strongly with a group, their behaviours tend to be consistent with norms of that group (social identity theory, \cite{tajfel2004social}). The literature on social influence discusses the role of embeddedness by assessing how peers' opinions, including both ``descriptive norms'' (the consciousness of what others do) and ``subjective/injunctive norms'' (the perception about what others believe), map into individual motivations and actions (\citealt{ajzen1991theory, cialdini1990focus}).

When referring to food, individual opinions and behaviours, and the influence that social norms exert on them, shows a unique complexity. The literature on consumer psychology has recognized that food choices result from a wide range of idiosyncratic factors, including moods, distraction, sensory cues, and psychology (\citealt{bublitz2010}), as well as from social influence (\citealt{mcferran2010}). Food decisions tend to be driven by habits even when consumers report an intention to do otherwise (\citealt{jiandwood2007,quested2013spaghetti}). Furthermore, people prefer to be in tune with others on the values and beliefs driving their food choices (\citealt{jackson2005economics}). In this regard, significant association has been found between subjective norms and individual opinions influencing food decisions (\citealt{graham2015predicting, stancu2016determinants}).

Habits tend to play a role in consumers' food waste behaviour too (\citealt{lewin1951field, verplanken1998habit, stern2000new, graham2014identifying}). While ethnographic evidence has shown that food waste behaviour includes a social component (\citealt{evans2011blaming}), the effect of habits tends to overcome personal intentions and to hinder social pressures even if consumers are averse to waste (\citealt{bolton2012less}). Furthermore, food waste stems from private behaviours (\citealt{quested2013spaghetti}); therefore, it is not immediately visible to peers (\citealt{cecere2014waste}) and no evidence of the relationship between descriptive norms and individual opinions emerge (\citealt{lapinski2005explication,graham2015predicting}). On the one hand, \cite{lamberton2013destination} show that consumers’ private decisions are grounded on one’s own ideas and perceptions, and are thus less affected by others' reasoning in social interactions. On the other hand, \cite{huh2014social} found that consumers are more likely to mimic others' decisions when choosing in private, especially if their own preferences are not well-defined.

Why would individuals change their food waste behaviours as a result of opinion interactions with peers? Literature suggests that persuasion (\citealt{jackson2005economics}), social stigma (\citealt{kerr2008detection,kurzban2001evolutionary}) and social punishment (\citealt{bowles2004evolution,dreber2008winners,fehr2005human,fowler2005altruistic,fowler2005egalitarian}) can be effective in steering individual choices toward cooperative (socially optimal) outcomes. Consumer food waste behaviour at home is private; therefore, such strategies cannot work. Nevertheless, when choices are influenced by strong habits as in the case of food, people tend to adapt their behaviours to those of their peers through a process of mimetism to minimize their perceived distance from them (\citealt{van2004mimicry}) and achieve consistency in their social relations (\citealt{jackson2005economics}). This process is not linear: in the context of food, attitudinal ambivalence, e.g. between enjoying large portions and avoiding waste, is common (\citealt{corniletal2014acuity}). \cite{piras2021community} have shown that food waste is a social dilemma between care for the society (which is more common in areas with high social capital measured as political participation and blood and organ donations) and individual status deriving from food abundance. An ``acceptable range of dishonesty'' (i.e. discordance between behaviours and opinions) exists within which consumers do not experience personal discomfort (\citealt{argo2012}), but if this range is overcome, they need to solve this contradiction by means of a cognitive mechanism (\citealt{festinger1962}).

The mismatch between publicly-stated opinions and private behaviours, also highlighted by the difference in food waste levels when detected through questionnaires and through waste sorting analysis (\citealt{giordano2018questionnaires}), represents a challenge for designing interventions aimed at tackling food waste. The literature has identified these dynamics at individual and social level, but to the best of our knowledge, their aggregate outcome in terms of behaviours has not been assessed. To explore how private food waste behaviours are affected by social exchanges of opinions in the presence of the above mismatch, we build an ABM based on the framework of opinion dynamics (\citealt{deffuant2002can,weisbuch2002meet, hegselmann2002opinion,lorenz2007continuous,galam2008sociophysics}). In this class of models, individual opinions evolve through progressive coupling, leading to the creation of clusters of individuals with similar opinions. We draw further inspiration from \cite{weisbuch2002meet} and \cite{deffuant2002can}, who study populations where individuals have \textit{bounded confidence}, i.e. are unwilling to compromise with individuals they consider too far from them. Namely, we include an opinion-distance threshold beyond which the agents are not updating their opinions in binary encounters. As recently shown by \cite{young2017can}, and debated by \cite{young2017social} and \cite{MattGavinAswer2017}, social media tools have some affect in reducing private food waste, but face-to-face interactions have a stronger potential to effectively modify behaviours. Hence, even if the interactions in our model could theoretically happen online, we developed it with face-to-face interactions in mind.

We extend the setup of \cite{weisbuch2002meet} by allowing agents' opinions and behaviours to co-evolve, and the structure of social networks to constrain their interaction patterns. Individuals are modelled as heterogeneous in their sociability, their initial opinions and behaviours concerning food waste, their willingness to change opinion when interacting with peers (\citealt{NishiMasuda2013}), and their tolerance for what we call ``cognitive dissonance'', i.e. the discordance between opinions and behaviours discussed above. For the sake of our analysis, simulated populations are divided into groups characterized by specific behavioural patterns. In each group, the variables are assigned either homogeneous values, or group-specific distributions. This modelling strategy allows to reduce unnecessary complexity (and thus complicatedness) by focusing on the group dychotomies which are of interest for the researcher, thus facilitating the identification of what variables drive the results. By means of extensive simulations, we study the impact on food waste behaviours of interactions between individuals belonging to different groups. After illustrating the baseline dynamic of the model by considering two theoretical groups, we calibrate it using empirical data and insights from the literature.

\section{Data and methods}\label{sec:method}
This section describes the theoretical model and the data used to characterize the two case studies through calibration.

\subsection{The theoretical model}
We consider a set of N agents $i\in\{1...,n\}$, representing single consumers (or, equivalently, households as unitary decision-makers\footnote{In the real world, members of the same household can see and pass judgements based on each other's food waste behaviour, which cannot be observed publicly. However, to the best of our knowledge, consumers' food waste has only been measured at aggregated household level. Therefore, we treat food waste behaviours as pertaining to a single, unitary agent, and intra-household dynamics are out of the scope of our model.}). Each agent has a time-evolving food waste behaviour $W_i^t\in[0;1]$ which measures the amount of waste generated in food routines (planning, purchasing, storing, preparing and cooking, serving and eating, disposing), with zero indicating the least wasteful behaviour. The behaviour of each agent is a private information, and is not observable by others. However, consumers exchange with their peers opinions on food waste $O_i^t\in[0;1]$, e.g. on its acceptability or their concerns about its social and environmental impacts, with zero indicating opinions most averse to food waste. 

Agents' interactions are bound by a network structure representing the ensemble of peers with which they exchange ideas and opinions (in line with \citealt{weisbuch2002meet}). Indeed, social influence does not spread linearly among consumers but the position of individuals in their networks and their relative influenceability (here characterised as ``openness of mind'') matter (\citealt{watts2007}). Formally, each agent $i$ has a set of peers $i' \in \mathcal{S}_i \subset (N \neq i)$, with reciprocal connections, i.e. $i \in \mathcal{S}_{i'} \Leftrightarrow i' \in \mathcal{S}_i$. We consider the social network linking the individuals as an Erd{\"o}s-R{\'e}nyi random network topology (\citealt{erdos1959random}) with density $\lambda\in[0;1]$. Random networks introduce a limited amount of variance in the network's degree, i.e. the number of connections of individual agents; however, as shown in Fig. \ref{FigS3} in Appendix, our results are robust to the elimination of such variance. We refrain from adopting complex network structures, where the presence of abnormally connected individuals may complicate the interpretation of the results. Furthermore, despite characterizing the populations by attributing different social and waste features to different groups, we assume no ex-ante connection between agents' initial food waste and the characteristics of their networks. Finally, the model was developed with face-to-face interactions in mind but nothing prevents these dyadic interactions from happening online or via other communication tool. Instead, one-to-many communication (e.g., information diffused via TV or online platforms) is not considered in our setting but could be introduced as an additional treatment.

Interacting individuals tend to end up with opinions more similar to their peers. However, in line with \cite{weisbuch2002meet}, we assume our agents to be unwilling to compromise with agents expressing opinions which are too far from their priors. We define $d_{int}^i\in[0;1]$ as the ``interaction threshold'', i.e. the distance beyond which others' opinions are not taken into consideration. It essentially measures the opposite of the intensity of the ``confirmation bias'' expressed by an individual. Interactions progress in discrete time steps, with one existing tie between two randomly-chosen agents $i$ and $i'$ being selected at each step $t$. Then, the opinion of agent $i$ evolves according to the rule: 
\begin{equation}
O_i^{t+1}= 
\begin{cases}
O_i^t + \mu (O_{i'}^t - O_i^t) & \text{ if } \,\,\,\,\, \vert O_{i'}^t- O_i^t \vert < d_{int}^i \\
O_i^t & \text{otherwise}
\end{cases}
\label{opdyn}
\end{equation}
where $\mu \in [0;0.5]$ is a parameter indicating the speed of convergence ($\mu=0.5$ in all simulations, meaning that a simple average between the opinions of the two agents is calculated; lower values would simply have the effect of slowing down the evolution of the model). The same applies to the interacting partner $i'$. However, note that in Eq. \ref{opdyn}, $d_{int}^i$ does not need to be equal to $d_{int}^{i'}$. Thus, this setup admits interactions where consumer $i$ changes opinion while consumer $i'$ does not, and vice versa, meaning social influence could be asymmetric. Like in the dyadic framework of decision-making proposed by \cite{simpson2012relat}, interacting consumers end up with intermediate opinions that are the result of reciprocal influence.

While agents' opinions start to evolve due to social interactions, waste behaviours (actions) do not necessarily evolve accordingly. Indeed, actions tend to be stickier than opinions. The fact that individual consumers declare a specific opinion when discussing with others (e.g., when reporting about their own food waste behaviour) while acting differently in private is a well-documented feature of food waste surveys (\citealt{quested2013spaghetti,moller2014standard,moller2014report,SettietAl2016,Vangeffen2016standard,giordano2018questionnaires}). \cite{sengupta2002misrep} find that misrepresentation is common in the context of social interactions among consumers when this is likely to generate a positive self-image, e.g. of an environmentally concerned person. This finding is especially relevant for food waste, whose avoidance carries strong ``social desirability'' (\citealt{giordano2018questionnaires}). However, if actions and opinions become too different compared to a given idiosyncratic threshold $d_{cd}^i$, i.e. when $\vert W_i^t - O_i^t \vert > d_{cd}^i$, agents start to suffer from cognitive dissonance, since they must defend in public an opinion they do not abide by in private. When this happens, the opinion $O_i^t$ and the action $W_i^t$ move closer. We define $I_i^t$ as the fraction of times an agent has modified their opinion (regardless of the size of the variation) over the number of past interactions, which is a proxy of their influenceability. If the number of past interaction is equal to zero, $I_i^t = 0.5$. When agents start to suffer from cognitive dissonance, their action or opinion change according to this rule:
\begin{equation}
\begin{cases}
O_i^t = (1 - I_i^t) O_i^t + I_i^t W_i^t & \text{with probability}  \,\,\,\,\, I_i^t \\
W_i^t = (1 - I_i^t) O_i^t + I_i^t W_i^t & \text{with probability}  \,\,\,\,\, 1-I_i^t \\
\end{cases}
\label{cddyn}
\end{equation}
This is in line with \cite{mukhopadhyay2008} who enquired consumers' reaction to food temptations in the form of resisting or succumbing, finding that individuals end up repeating past decisions in the present.\footnote{Let us note that, on the top of ``cognitive dissonance'', the society as a whole could be affected by a systematic ``desirability bias'' that applies to food waste actions. If this is the case, however, such bias would not qualitatively change the results of the model, but rather just imply a systematic shift (a bias) in these results.}

Although social interactions tend to remain stable over time, when faced with an opinion that is too distant from their private action, individuals could feel uncomfortable and thus reduce the contacts with the person expressing it. Formally, an agent $i$ considers whether $\vert W_i^t - O_{i'}^t \vert < d_{act}^i$, and if this condition is not satisfied, they eliminate the tie and create a new one.\footnote{In the simulations reported in this paper, $d_{act}^i=0.1 \,\,\, \forall i$; preliminary simulations showed that small variations around this value do not significantly change our results.} In line with the classical sociological literature (\citealt{simmel1908sociology}), and specifically with the observation that most new ties are created with friends of friends (triadic closure), the new link is created with an individual already connected with one's friends. While this rule is followed in most cases, there is a small probability $P_{rand}<<1$ that the new link is formed randomly.\footnote{In the simulations presented in this paper $P_{rand}=0.01$; preliminary simulations showed that small variations around this value do not change the outcome.}

The parameters and variables in the model, their symbols, and their admitted range of values are illustrated in Table \ref{ModelParameters}. For parameters, also the fixed values is thereby included.
\begin{table}
\centering
\caption{Parameters and variables used in the model, symbols, and admitted values.}
\scalebox{0.8}{
\begin{tabular}{p{9cm}  c c  p{3cm}  } 
\toprule
\textbf{Parameter} & \textbf{Symbol} & \textbf{Possible values} & {\bf Parameter values $\forall$ simulations} \\ 
\midrule
Max number of time steps (dyadic encounters) & $t_{max}$ & $[1; \infty]$ & $4\cdot10^{5}$ \\ \hline
Number of agents in the population & $N$ & $[2; \infty]$ & 1,000 \\ \hline
Food waste action of each agent $i$ at $t_0$ & $W_i^0$ & $[0;1]$ \\ \hline
Food waste opinion of each agent $i$ at $t_0$ & $O_i^0$ & $[0;1]$ \\ \hline
Half-range of the distributions of $W_i^0$ and $O_i^0$ (for case study calibration) &  $\sigma$ & 
$[0;0.5]$ \\ \hline
Density of the network linking the agents & $\lambda$ & $[0;1]$ & $0.5$ \\ \hline
Interaction threshold (distance beyond which others' opinions are not taken into consideration) & $d_{int}^i$ & $[0;1]$ \\ \hline
Speed of convergence of the opinions (weight in the averaging mechanism) & $\mu$ & $[0;1]$ & $0.5$ \\ \hline
Cognitive dissonance (max tolerated distance between $W_i$ and $O_i$) & $d_{cd}^i$ & 
$[0;1]$ \\ \hline
Weight in the movement of $W_i$ towards $O_i$ in case of no previous interactions & $I_i^0$ & $[0;1]$& $0.5$ \\ \hline
Distance $W_i^t - O_{i'}^t$ over which a tie is eliminated & $d_{act}^i$ & $[0;1]$ & $0.1$ \\ \hline
Probability that a new tie is formed randomly & $P_{rand}$ & $[0;1]$ & $0.01$ \\ \hline
Probability that an agent of group $j$ is connected with another individual from the same group $j$ & $P_{intra}$ & $[0;1]$ & \\
\bottomrule
\end{tabular}
}

\label{ModelParameters}
\end{table}
The model described in this section as well as all simulations are implemented in MATLAB 2016.\footnote{All source codes used to create the figures and the data generated by the simulations are available at the following link: \url{https://github.com/simonerighi/FoodWasteModel}.}

\subsection{Model calibration}
The model described above is purely theoretical, i.e. no specific values are assigned to its parameters. To demonstrate the model's helpfulness in studying the evolution of food waste behaviours in a real population, we calibrate it using data on household food waste in Italy. We choose Italy because the model was developed at the University of Bologna, and because of the lack of evidence about the effectiveness of campaigns, and of large-scale simulations of food waste levels in this country. Italy represents an interesting case study due to the wide economic and socio-political diversity across its territory, which has been extensively studied in the literature (\citealt{putnam1993making}), and which results in different levels of social commitment towards environmental goals, including food waste reduction (\citealt{piras2021community}).

For calibrating the model, we use secondary data collected by the Italian National Observatory ``Waste Watcher''\footnote{Waste Watcher: \url{http://www.sprecozero.it/waste-watcher/} [Accessed 11.02.2021].} in 2015. ``Waste Watcher'' was established by Last Minute Market,\footnote{Last Minute Market: \url{http://www.lastminutemarket.it/} [Accessed 11.02.2021].} a spin-off of the University of Bologna that works towards food waste reduction through the recovery and redistribution of unsold food products. It has been running yearly surveys on consumer food waste since 2013. Although consumer food waste in Italy has been assessed through a number of one-time surveys (\citealt{gaiani2018food}), this is the longest available time series; this long-term experience allowed to progressively refine the questionnaire. The survey is administered via CAWI (Computer-Assisted Web Interviewing) by the market research company SWG.\footnote{SWG: \url{https://www.swg.it/} [Accessed 11.02.2021].} To achieve representativeness of the Italian population, quota sampling is adopted, with stratification by age, sex, income, and geographical area. Consumers belonging to a pool owned by the company are contacted via email, and asked to fill the questionnaire online until reaching the quotas. The 2015 survey involved a sample of 1,502 individuals aged 18 and above. Respondents had to be the responsible person, or one of the responsible persons, for food shopping, management or cooking within the household.\footnote{The questionnaire consisted of around 90 closed-ended questions (including conditional ones) on socio-demographic characteristics (about 30 questions), food routines, opinions on food waste or related issues such as the resources used to grow food, and policy interventions to prevent food waste. Two of the authors of this paper contributed to the development of some of the questions. Filling the questionnaire took 20-25 minutes on average. The dataset was provided to the authors by Last Minute Market. A subset of the questionnaire including all the questions used for calibration is provided as Supplementary Material. For research purposes, the full questionnaire can be requested to Last Minute Market.}

In the ``Waste Watcher'' dataset, food waste behaviours are self-assessed and self-reported. This is done using three ordinal-scale questions on food waste frequency, quantity, and monetary value.\footnote{The questions and potential responses are provided as Supplementary Material. The frequency is measured on a five-point scale, the quantity on a six-point scale, and the value on a nine-point scale.} Despite slight differences,\footnote{For example, in 2013 the EU-level Eurobarometer survey asked about the \textit{percentage} of food purchased that ended up as waste (\citealt{grainger2018use}).} ordinal-scale questions are the norm in questionnaires to quantify food waste (\citealt{gaiani2018food,giordano2018questionnaires}). \cite{giordano2018questionnaires} found that self-assessed food waste levels are subject to underestimation. However, in most countries household food waste has been estimated exclusively through questionnaires due to the lower cost of this methodology, and this was the only available estimate of household food waste in Italy when we developed the model (\citealt{gaiani2018food}). Food waste opinions are detected by means of three ordinal-scale questions on the respondent's perception of the scale of the food waste problem, its seriousness, and how much they worry about it.\footnote{The questions and potential responses are provided as Supplementary Material. The scale is measured on a three-point scale, the seriousness on a four-point scale, and the worrying on a four-point scale.} Although the main themes addressed by the survey remained the same along the years, specific questions, including those to assess food waste, have been refined based on new literature findings, projects, and empirical evidence, thus improving their effectiveness in detecting true population values.

Since opinions and behaviour of different individuals within a household could be heterogeneous, representing households with the response of a single member leads to reduced accuracy. Nevertheless, only a few studies have attempted to collect and analyse data on environmental behaviours from household members separately. For example, \cite{longhi2013individual} found that in the UK there are disparities in pro-environmental values and behaviour between household members when controlling for characteristics such as gender, age, education, and employment status. \cite{seebauer2017disentangling} reached similar results after interviewing live-in couples in Austria. To the best of our knowledge, the individual vs household issue has never been investigated in the food waste literature where instead household size has been identified as one of the main determinants of waste levels (\citealp{grainger2018use,grainger2018model, SettietAl2016, stancu2016determinants, parizeau2015household, secondi2015household, koivupuro2012influence, barr2007factors}). Indeed, due to the nature of the food waste phenomenon, individual waste flows are undistinguishable with any set of data, meaning that for all intent and purposes, the households need to be considered as unitary decision-makers. Also the ``Waste Watcher'' survey asks about food waste behaviours at the level of household but then measures the opinions of the individual respondent, implying that the latter is representative of the entire household and thus treating the household as a ``black box''. Equally, in our model the agents can be understood as households with a single aggregated food waste opinion and a single aggregated food waste behaviour, and whose links with other individuals are the sum of the links of all the household members. This assumption does not change the nature of the model, whose findings in terms of impact of social interactions remain valid. Entering intra-household dynamics leading to waste could be an interesting venue for future research, including modelling efforts.

While individuals are in principle all different from each other, we took steps to organize consumer differences around archetypal socio-demographic groups. Despite some degree of within-group variation, it is reasonable to hypothesise intra-group heterogeneity to be smaller than inter-group heterogeneity. Therefore, we assume out the former, and focus on the impact of the latter. Some parameters, like the interaction thresholds $d_{int}^i$, are set as homogeneous within each group $j$: 
\begin{equation}
\forall i\in j \,\,\,d_{int}^i=d_{int}^{j}. \\
\label{neweq}
\end{equation}
Instead, we assume that the initial food waste actions $W_0$ and opinions $O_0$ of the individuals belonging to a group $j$ follow a distribution specific to that group. In particular, these variables are calibrated as follows. For each groups $j$, the probability mass function of the three behaviours and the three opinions is determined from the dataset; the 0--1 range is divided into as many classes as the options in each question; for each agent $i$, and each of the three behaviours and the three opinions, a value in the 0--1 range is extracted using the probability mass function; finally, the three behaviours and the three opinions are averaged to obtain respectively the initial food waste action $W_0^i$ and opinion $O_0^i$. The idiosyncratic ``acceptable range of dishonesty'' is set equal to the initial distance between the action and the opinion $d_{cd}^i$=\(\lvert W_i^0-O_i^0\rvert\). This procedure implies a number of subjective choices. However, the purpose of this calibration is to set these key variables so that their distribution and correlation are qualitatively realistic rather than to achieve full representativeness of household food waste levels in Italy.

Differently from food waste, the values assigned to the parameters illustrating socio-psychological constructs ($\lambda$, $d_{int}^i$, $\mu$, $d_{cd}^i$, $I_i^0$, $d_{act}^i$, $P_{rand}$ and $P_{intra}$) are purely theoretical. To the best of our knowledge, no studies in the food and food waste realm have ever tried to quantify them. For this reason, rather than exploring reality in its full complexity, for some of them we use a simple parametrisation described in the previous subsection. For others, we explore the impact on our output variable of changing their value endogenously through numerical comparative statics. This is a common practice in the ABM literature. If in the future these parameters will be measured by means of surveys, it will be possible to calibrate them too.

\paragraph{Identification of the case studies.} As a first step, we identified homogeneous groups and their relative size within the sample. The dychotomy between young and adult people, with the former wasting more food, has been recognised as particularly relevant in the literature (\citealt{quested2013spaghetti, stefan2013avoiding, watson2013food, Wrap2014a, stancu2016determinants}). For this reason, our first case study focuses on this dychotomy, rather than on other ones such as education or income.\footnote{Deriving more than two groups by cross-tabulating several variables would have resulted in small sub-samples, reducing representativeness and complicating the interpretability of the results in terms of mixing.} This case study is particularly relevant for Italy due to the strength of families and the importance of intergenerational contacts compared to other countries (e.g., the role of grandparents in children’s education, \citealt{albertini2016ageing}). Among others, this could favour the transmission of food management practices. For this first case study, survey respondents are classified as adults if aged 45 years or more (58\% of the weighted sample), as young if aged between 18 and 44 (42\%). The choice of this threshold is a pragmatic one, aimed at generating two groups of similar size and deriving two conditional distributions; other thresholds could have been used without affecting the validity of the results.

\cite{Zemborain2007} suggest that ambivalent individuals, who do not hold a clear opinion on an issue, are more open to persuasion. Therefore, a second case study is constructed around the dychotomy between committed and ordinary (non-committed) individuals. This is again a relevant case study for Italy, which has been characterised since its unification by a ``silent majority" of moderate individuals (\citealt{gramsci1975quaderni}), and small elite groups extremely committed to universal value (thus, potentially, also environmentalism). Committed respondents are identified as those who selected the lowest available option in all the six questions on food waste behaviours and opinions. Overall, they account for 5.4\% of the population in the ``Waste watcher'' sample. A Wilcoxon rank-sum test was used to assess if the variables of interest differ significantly across groups.\footnote{This type of test is well-suited for ordinal data like most variables in our dataset, and is less sensible than a t-test to the presence of outliers in the skewed distributions of food waste opinions and behaviours.}

\paragraph{Young vs adult consumers.} In the ``Waste watcher'' sample, adults waste significantly less food both in terms of quantity and monetary value, and less frequently than the youth. After reporting these behaviours on a 0--1 scale and averaging them as above, the mean resulting action is 0.18 for the youth and 0.12 for adults. Adult respondents are also significantly more concerned about food waste, more likely to think that the amount of food waste at societal level is large, and to deem it a serious problem for the planet. After applying the same transformation as above, the value of the opinion becomes 0.33 for adults and 0.37 for the youth. The average ``range of dishonesty'' is only slightly larger for adults, the difference being probably due to stronger habits; indeed, rather than being driven by an intention not to waste food, adults' behaviours are embedded in food activities, including planning and shopping routines (\citealt{stefan2013avoiding}). Descriptive statistics are provided in Table \ref{YoungOldTableData}.

\begin{table}
\centering
\caption{Food waste indicators for young and adult respondents.}
\scalebox{0.8}{
\begin{tabular}{p{9cm} r r r}
\hline
\textbf{Variables} & \textbf{Youth} & \textbf{Adults} & \textbf{p-value (Wilcoxon)} \\ \hline
Share of households in the sample & 
0.4185 & 
0.5815 & 
- \\ \hline
Food waste frequency (times per day) & 
0.1212 & 
0.0769 & 
0.0000 \\ \hline
Food waste quantity (kilograms per week) & 
0.1895 & 
0.1258 & 
0.0000 \\ \hline
Food waste value (Euros per week) & 
7.24 & 
5.14 & 
0.0000 \\ \hline
\textit{Food waste action} (0=low; 1=high) & 
0.1813 & 
0.1235 & 
0.0000 \\ \hline
Perception of food waste scale (1=big; 3=small) & 
2.0675 & 
1.9937 & 
0.0012 \\ \hline
Seriousness of food waste for the planet (1=very serious; 4=not serious at all) & 
1.7206 & 
1.6148 & 
0.0092 \\ \hline
Worrying for food waste (1=a lot; 4=not at all) & 
1.9770 & 
1.8840 & 
0.0223 \\ \hline
\textit{Food waste opinion} (0=averse; 1=favourable) & 
0.3665 & 
0.3321 & 
0.0000 \\ \hline
Cognitive dissonance (0=action equal to opinion; 1=maximum distance) & 
0.2467 & 
0.2428 & 
0.0665 \\
\bottomrule
\end{tabular}
}
\label{YoungOldTableData}
\end{table}

In the simulations, the values of $W_0^i$, $O_0^i$ and $d_{cd}^i$ are assigned to each individual on the basis of data; instead, the interaction threshold $d_{int}^i$ is varied for young people to assess the sensitiveness of the model and thus of food waste to its change. Table \ref{YoungOldParameters} provides an overview of the calibration parameters used in this first case study.

\begin{table}
\centering
\caption{Parameters used for calibrating the ``young vs adults'' case study.}
\scalebox{0.8}{
\begin{threeparttable}
\begin{tabular}{p{2cm} l l l} 
\toprule
\textbf{Variables} & \textbf{Youth} & \textbf{Adults} & \textbf{Origin} \\ 
\midrule
$t_{max}$ & \multicolumn{2}{ c }{$4\cdot10^5$} & a priori \\ \hline
$N$ & 418 &  582 & from data \\ \hline
$W_i^0$ & 0.1813 & 0.1235 & from data \\ \hline
$O_i^0$ & 0.3665 & 0.3321 & from data \\ \hline
$\lambda$ & \multicolumn{2}{ c }{0.5} & theoretical (see text) \\ \hline
$d_{int}^i$ & $[0.01;0.16]^*$ & $0.06 \,\,\, \forall i \in A$ & tested endogenously \\ \hline
$\mu$ & \multicolumn{2}{ c }{0.5} &  theoretical (see text) \\ \hline
$d_{cd}^i$ & \multicolumn{2}{ c }{\(\lvert W_i^0-O_i^0\rvert\) } &  from data \\ \hline
$I_i^0$ &  \multicolumn{2}{ c }{$0.5 \,\,\, \forall i$} &  theoretical (see text) \\ \hline
$d_{act}^i$ &  \multicolumn{2}{ c }{$0.1 \,\,\, \forall i$} &  theoretical (see text) \\ \hline
$P_{rand}$ &  \multicolumn{2}{ c }{0.01} &  theoretical (see text) \\ \hline
$P_{intra}$ &  \multicolumn{2}{ c }{$[0;1]^{**}$} & tested endogenously \\ 
\bottomrule
\end{tabular}
\begin{tablenotes} 
{\footnotesize \textbf{Notes:} *in steps of 0.05 ** in steps of 0.025 \par}
\end{tablenotes}
\end{threeparttable}
}
\label{YoungOldParameters}
\end{table}

\paragraph{Committed vs ordinary consumers.} In the ``Waste watcher'' sample, committed individuals waste significantly less food in terms of both quantity and monetary value, and less frequently than ordinary individuals by definition. As a result, their average food waste action is zero, compared to 0.16 for ordinary people. Equally, committed respondents are more likely to be concerned about food waste, to think that the amount of food waste at societal level is large, and to consider it a serious problem for the planet. Their average food waste opinion is zero by definition, compared to 0.37 for ordinary people. The average ``acceptable range of dishonesty'' of committed individuals is zero by construction, but the difference with ordinary people is not statistically significant. Descriptive statistics are provided in Table \ref{committedData}.

\linespread{1.0}
\begin{table}
\caption{Food waste indicators for committed and ordinary respondents.}
\centering
\scalebox{0.8}{
\begin{tabular}{p{9cm} c c p{2cm}}
\hline
\textbf{Variables} & \textbf{Committed} & \textbf{Ordinary} & \textbf{p-value (Wilcoxon)} \\ \hline
Share of households in the sample & 
0.0536 & 
0.9464 & 
- \\ \hline
Food waste frequency (times per day) & 
0.0000 & 
0.1008 & 
0.0000 \\ \hline
Food waste quantity (kilograms per week) & 
0.0000 & 
0.1611 & 
0.0000 \\ \hline
Food waste value (Euros per week) & 
2.50 & 
6.22 & 
0.0000 \\ \hline
\textit{Food waste action} (0=low; 1=high) & 
0.0000 & 
0.1560 & 
0.0000 \\ \hline
Perception of food waste scale (1=big; 3=small) & 
1.0000 & 
2.0826 & 
0.0000 \\ \hline
Seriousness of food waste for the planet (1=very serious; 4=not serious at all) & 
1.0000 & 
1.6963 & 
0.0000 \\ \hline
Worrying for food waste (1=a lot; 4=not at all) & 
1.0000 & 
1.9752 & 
0.0000 \\ \hline
\textit{Food waste opinion} (0=averse; 1=favourable) & 
0.0000 & 
0.3662 & 
0.0000 \\ \hline
Cognitive dissonance (0=action equal to opinion; 1=maximum distance) & 
0.0000 & 
0.2582 & 
0.6206 \\
\bottomrule
\end{tabular}
}
\label{committedData}
\end{table}
\linespread{2.0}

Also for this second case study, in the simulations the values of $W_0^i$, $O_0^i$ and $d_{cd}^i$ are assigned to each agent on the basis of data, while the interaction threshold $d_{int}^i$ is varied for committed individuals to assess its impact on food waste levels. Even if the variability of $W_0^C$ and $O_0^C$ is zero by construction, to implement our sensitivity analysis we allow for some intra-group heterogeneity. After assigning the values of $W_0^i$ and $O_0^i$ as described above, their values for committed agents are re-extracted from uniform distributions centred around $\overline{W_0^{C}}$ and $\overline{O_0^{C}}$ respectively, and with half-range $\sigma^{C}\in[0;0.2]$. The impact of $\sigma^{C}$ is tested in the model. The values of $W_0^i$ and $O_0^i$ for ordinary individuals are kept as extracted initially. Table \ref{committedParameters} provides an overview of the calibration parameters used in this second case study.

\begin{table}
\centering
\caption{Parameters used for calibrating the ``committed vs ordinary'' case study.}
\scalebox{0.8}{
\begin{threeparttable}
\begin{tabular}{p{2cm} c c l} 
\toprule
\textbf{Variables} & \textbf{Committed} & \textbf{Ordinary} & \textbf{Origin} \\ 
\midrule
$t_{max}$ & \multicolumn{2}{ c }{$4\cdot10^5$ }& a priori \\ \hline
$N$ &  54 &  946 & from data \\ \hline
$W_i^0$ & $\sigma^{C}\in[0;0.2]^{*}$ &  0.1560 &  tested endogenously; from data \\ \hline
$O_i^0$ & $\sigma^{C}\in[0;0.2]^{*}$  & 0.3662 & tested endogenously; from data \\ \hline
$\lambda$ & \multicolumn{2}{ c }{$0.5$ }&  theoretical (see text)  \\ \hline
$d_{int}^i$ & $ [0;0.05]^{**} \,\,\, \forall i \in C$ & $0.1 \,\,\, \forall i \in Ord$ &  tested endogenously \\ \hline
$\mu$ &  \multicolumn{2}{ c }{$0.5$ }&  theoretical (see text) \\ \hline
$d_{cd}^i$ & \multicolumn{2}{ c }{\(\lvert W_i^0-O_i^0\rvert\) } &  from data \\ \hline
$I_i^0$ & \multicolumn{2}{ c }{$0.5 \,\,\, \forall i$} & theoretical (see text) \\ \hline
$d_{act}^i$ & \multicolumn{2}{ c }{$0.1 \,\,\, \forall i$} & theoretical (see text) \\ \hline
$P_{rand}$ & \multicolumn{2}{ c }{$0.1$} & theoretical (see text) \\ \hline
$P_{intra}$ & \multicolumn{2}{ c }{$0.5$} & theoretical (see text) \\ 
\bottomrule
\end{tabular}
\begin{tablenotes} 
{\footnotesize \textbf{Notes:} *in steps of $0.0025$. **in steps of $0.005$. \par}
\end{tablenotes}
\end{threeparttable}
}
\label{committedParameters}
\end{table}

\paragraph{Interactions between groups.} The members of a social group may (and in general are expected to) have different interaction patterns with peers from their group and individuals from other groups. Obtaining data on the actual patterns of interaction across groups is exceedingly difficult, but since the networks of the two pairs of groups considered in the simulations may not be fully mixed, we explore different degrees of intra- and inter-group influence. After dividing the population into two groups with their relative characteristic degrees, i.e. the average number of links of each agent $\lambda_j$ and $\lambda_{j'}$ ($\lambda_j$=$\lambda_{j'}$ in all subsequent simulations; indeed, as noted in Fig. \ref{FigS2} in Appendix, changing the relative level of $\lambda_j$ and $\lambda_{j'}$ does not change the outcomes significantly), we model this feature by building the social networks according to the procedure hereby described. An individual belonging to group $j$ is connected with another individual from the same group $j$ with probability $P_{intra}$, and with individuals from the other group $j'$ with probability $1-P_{intra}$. When $P_{intra}=0$, every individual is linked only with members of the other group; oppositely, when $P_{intra}=1$, the two groups are completely separated. A fully mixed network is obtained when $P_{intra}=0.5$.

\section{Results and Discussion}\label{sec:result}
\begin{figure}[h!]
\includegraphics[width=1\linewidth]{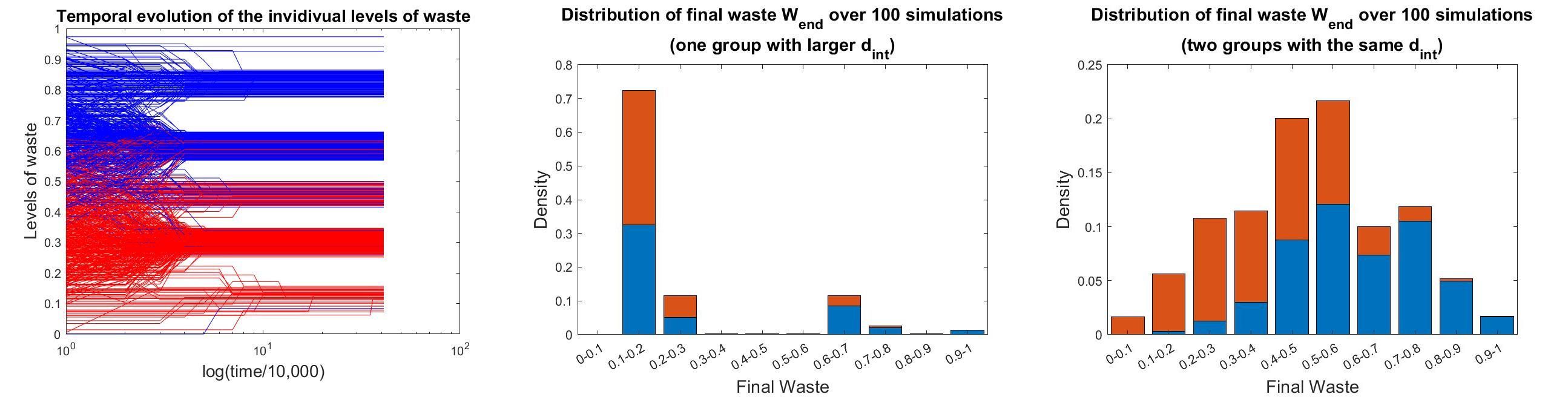}
\caption{\textbf{Dynamic behaviour of the model.} Left panel: temporal evolution of $W_i$'s in a representative simulation. Central panel: distribution of $W_i^{t_{max}}$ across 100 simulations (the interaction threshold is $d_{int}^R=0.10$ for red individuals and $d_{int}^B=0.20$ for blue individuals). Right panel: distribution of $W_i^{t_{max}}$ across 100 simulations with interaction threshold $d_{int}^B=d_{int}^R=0.10$. \textit{Notes}: In all panels $t_{max}=4\cdot10^{5}$, $P_{intra}=0.5$, $d_{cd}=0.05$, $\overline{W^R}=0.33$, $\overline{W^B}=0.66$, and $\sigma=0.33$ for both groups. $N=1,000$ agents, equally divided between the two groups.}
\label{Fig1}
\end{figure}

As a preliminary step, we consider the dynamic behaviour of our model. Fig. \ref{Fig1} shows the results of baseline simulations where two equally-sized groups of agents (blue and red) are characterized by the same level of confirmation bias $d_{int}^i$, and the same level of cognitive dissonance $d_{cd}$. The network density $\lambda$ is equal to 0.5. In both groups, the initial levels of waste follow a triangular distribution, with $Tri(\overline{W} \pm 0.33)$; however, the groups are heterogeneous in their initial average level of waste (0.33 for red consumers, 0.67 for blue ones). The left panel shows the evolution of the waste levels of each agent over time. Almost all the movements happen within the first 10,000 time-steps, especially among high-waste (blue) agents; then, the situation stabilizes, with clearly defined clusters emerging. Rather than converging towards an average level of waste, consumers tend to gather into four main equally-distanced clusters, and a smaller lower cluster; only a few individuals with very high waste remain isolated. Due to the perfect symmetry of our assumptions in this example all clusters are dominated by the type of consumers whose initial waste was closer to their centroid. However, the clusters located closer to the centre show a more diversified composition in terms of red and blue agents. By averaging the final waste of consumers across 100 simulations (central and right panels), it emerges that the final distribution of the agents in terms of waste depends on the relative openness of mind of the two groups. Allowing heterogeneity in the interaction threshold (i.e., if blue individuals, who waste more, are also more open-minded: right panel), the distribution becomes bi-modal, with a higher peak around low waste levels and a lower peak around mid-to-high waste levels. This happens because more open-minded (blue) individuals are more willing to compromise with others' opinions and thus tend to be convinced, and to move towards the position of less open-minded agents (red). The outcome changes sharply if the interaction threshold is the same for the two groups: the final distribution of waste resembles a bell-shaped distribution centered around 0.5, with a smaller peak around 0.7--0.8.

The emerging dynamics, showing multiple clusters rather than overall convergence of waste behaviours, finds a confirmation in reality but is not obvious from the modelling point of view. We replicate the findings of \cite{weisbuch2002meet}, whose model sees the emerging of multiple clusters ``from an initial configuration where transitivity of opinion propagation [is] possible through the entire population: any two agents however different in opinions could [be] related through a chain of agents with closer opinions'' \cite[p. 57]{weisbuch2002meet}. Such dynamics ``end up in gathering opinions in clusters on the one hand, but also in separating the clusters in such a way that agents in different clusters do not exchange anymore" \cite[p. 57]{weisbuch2002meet}. If such situation materialises, further improvement in terms of food waste levels through opinion exchange becomes very difficult; therefore, ``cognitive dissonance'' could hinder the effectiveness of waste reduction efforts. Although a certain convergence is observed in the central clusters, in line with the social identity theory (\citealt{tajfel2004social}), behaviours tend to align mostly with individuals from the same group, confirming that food waste has a social component (\citealt{evans2011blaming}).

\subsection{Adults vs young consumers}
\begin{figure}
\centering
\includegraphics[width=1\linewidth]{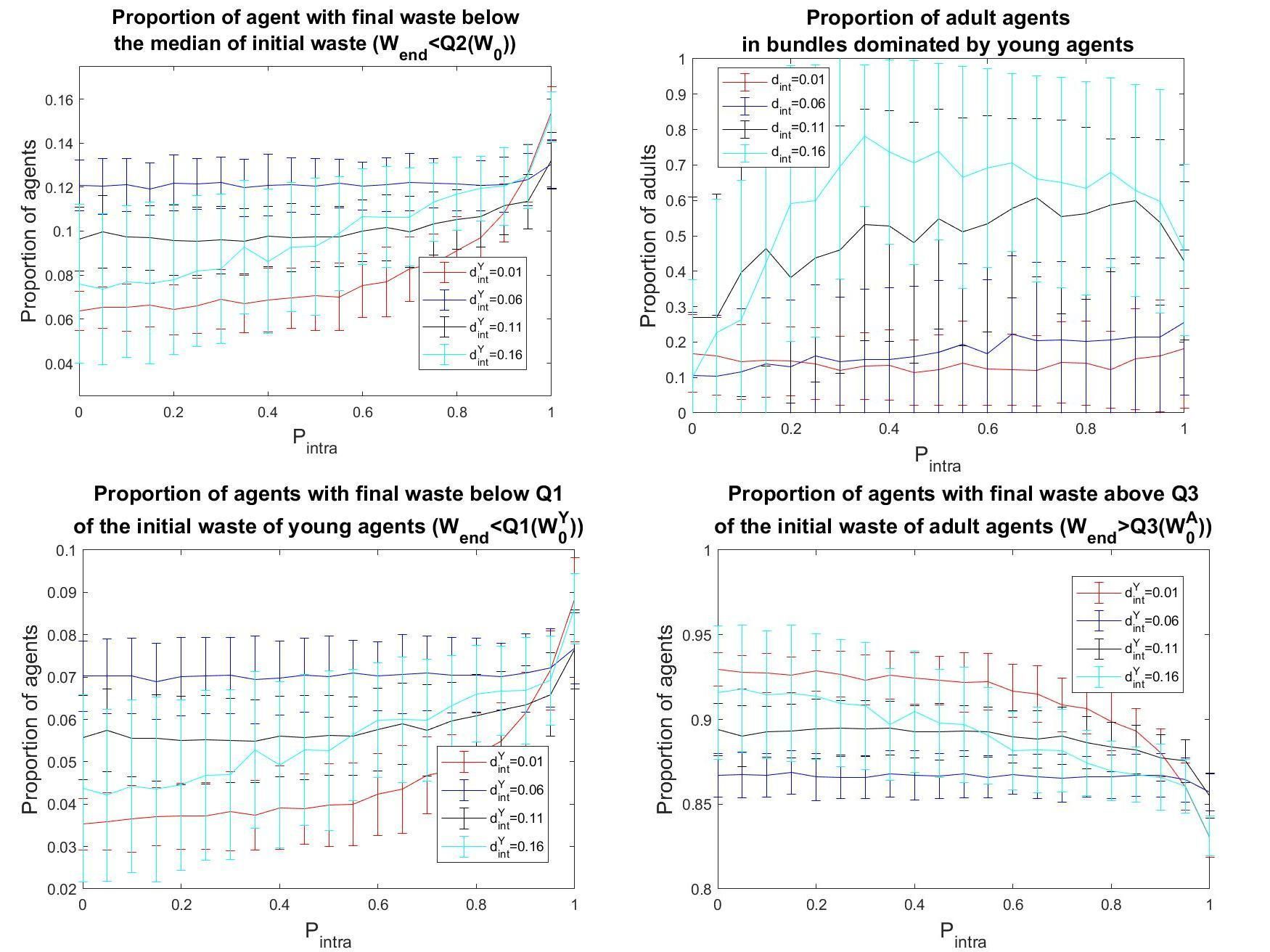}
\caption{\textbf{Adult vs young consumers.} Top-left panel: proportion of agents with $W_{end} < median(W_0)$. Top-right panel: proportion of adult agents ending up in bundles dominated (see definition in the text) by young agents. Bottom-left panel: proportion of agents with $W_{end} < Q_1(W_0^Y)$. Bottom-right panel: proportion of agents with $W_{end} > Q_3(W_0^A)$. \textit{Notes}: The calibration parameters are set in line with Table \ref{YoungOldParameters}. All the panels report the averages and standard deviations for changing levels of $P_{intra}\in[0;1]$ across 100 simulations. The interaction threshold for adults is fixed at $d_{int}^A=0.06$, while we explore $d_{int}^Y=\{0.01, 0.06, 0.11, 0.16\}$.}
\label{Fig2}
\end{figure}

Having shown the dynamic evolution of the model and its dependency on the main parameters, we now assign them the values illustrated in Table \ref{YoungOldParameters} to simulate the young vs adults dychotomy, and focus on the stable states reached in the long run. This allows to understand how the two groups influence each other's opinions and actions, shading light on the conditions in which social processes drive the reduction of private anti-environmental behaviours. In line with our calibration, our populations include 58\% of adult and 42\% of young agents, and adults present lower waste levels ($\overline{W_0^{A}}$=0.124) compared to young agents ($\overline{W_0^{A}}$=0.181), which is also in line with the literature (\citealt{stancu2016determinants}). The ``acceptable range of dishonesty'' $d_{cd}^i$, and thus the likelihood that a change in opinion is translated into a change in behaviour, is set for each agent as equal to \(\lvert W_i^0-O_i^0\rvert\), and does not differ significantly between the two groups.
Since it is difficult to pinpoint the sign of the relationship between the confirmation bias of the two groups (which is largely dependent on individual and situational factors), we explore different cases by means of simulations. The interaction threshold for adults is thus fixed at $d_{int}^A=0.06$, while for young agents it varies between $d_{int}^Y=0.01$ and $d_{int}^Y=0.16$. This domain allows to explore situations where adults are both more and less open-minded than young consumers. Furthermore, as the groups of young and adults are unlikely to be fully mixed, we explore different patterns of intra- and inter-group mixing by changing $P_{intra}$. For each combination of $d_{int}^Y$ and $P_{intra}$, Fig. \ref{Fig2} shows the average and the standard deviation of the waste levels obtained with 100 simulations, while an analysis of the sensitivity of the variables plotted to the values of $d_{int}^i$ and $P_{intra}$ is provided as Supplementary Material. As noted in the baseline scenario, open-minded individuals tend to be attracted in the direction of less open-minded ones through progressive coupling. The analysis is focused on the relationship between the strength of this attraction and the strength of the interaction among groups. 

Our sensitivity analysis shows that the relation between each of the variables plotted and the values of $d_{int}^Y$ and $P_{intra}$ is statistically significant regardless of the statistical model used to assess it. A key stylised fact emerging from Fig. \ref{Fig2} is that a large share of consumers tend to increase their waste levels, with more than 80\% ending up with a waste above the third quartile of the initial waste of adult agents, and less than 15\% with a waste below the initial median of the population. These large movements confirm that consumers can align their behaviour to what they think other people do (based on the latter's opinions) even when deciding in private (\citealt{huh2014social}). The increase in waste levels happens regardless of the initial setting, and is a result of calibration. Since in the ``Waste watcher'' dataset the opinions are much ``higher'' than the declared waste levels, they tend to attract the latter. A second relevant dynamic is represented by the interplay between the degree of mixing $P_{intra}$ and the interaction threshold $d_{int}^i$. If the interaction threshold differs between the two groups, small degrees of mixing ($0.8<P_{intra}<1$, i.e. when less than 20\% of the links are with agents from the other group) are enough to achieve sizeable movements, because less open-minded agents attract the members of the other group towards their position (in turn, the sensitivity analysis shows that values of $P_{intra}<0.25$ make a very small or non-significant difference compared to $P_{intra}=0.$). If the two groups present the same interaction threshold, the influence is reciprocal and the outcome in terms of waste levels does not depend on the level of mixing. Equally, increases in the level of mixing beyond 20\% do not make a large difference for the final outcome for given levels of $d_{int}^Y$.

In case of no mixing between the groups, around 85\% of the agents end up with a final waste above $Q_3(W_0^A)$ (bottom-right panel of Fig. \ref{Fig2}), around 14\% with a waste below $median(W_0)$ (top-left panel of Fig. \ref{Fig2}), and only around 8\% with a waste below the $Q_1(W_0^Y)$ (bottom-left panel of Fig. \ref{Fig2}). In a fully mixed population ($P_{intra}=0.5$), the share of agents with a waste above $Q_3(W_0^A)$ varies between about 86\% and 93\%, while between 7\% and 12\% end up below $median(W_0)$, and between 3\% and 7\% below $Q_1(W_0^Y)$. Given the generalised increase in waste levels, rather than the absolute changes it is more interesting to discuss the changes \textit{comparatively} for different parameter settings. The highest waste levels are observed when $d_{int}^Y < d_{int}^A$ and young agents are thus less willing to compromise with less wasteful ones (red lines). Interestingly, the lowest waste levels are not achieved when the interaction threshold of the young is much higher than the one of the adults ($d_{int}^{Y}=0.16$, light blue lines), but when the two groups present similar interaction thresholds ($d_{int}^{Y}=0.06$, blue lines). This is because large interaction thresholds imply contrasting influences ending up in less movement, as confirmed by the standard deviation of the final waste levels, which is much larger when the agents present a large interaction threshold (light blue bars), especially for higher levels of mixing ($P_{intra}>0.5$). Finally, the top-right panel of Fig. \ref{Fig2} suggests that young consumers are more likely to dominate a bundle\footnote{A bundle is characterized as a set of consumers less distanced among themselves than $max(P_{cd},P_{int})$, a radius equal to the maximum between the interaction threshold and the ``cognitive dissonance'' threshold. Such a group cannot further converge because the cognitive dissonance mechanism is never triggered and their interactions at the level of opinions do not result in a change in actions. This is similar to the mechanism that causes the emerging of non-converging clusters in \cite{weisbuch2002meet}. The dominance of a group of agents within a bundle is assessed by considering their relative share within that bundle. A bundle of consumers at $t_{max}$ is classified as dominated by agents of type $j$ if their proportion in the group is larger than their proportion in the population plus two standard deviations (computed relative to the size of the group).} when they present a \text{larger} interaction threshold and the degree of mixing is neither too low nor too high ($0.2<P_{intra}<0.8$). The graph does not allow to determine the reason of this outcome. However, it is likely that young agents move towards adults until becoming a majority in many adults' groups, while adults make smaller movements towards the young (as shown by the increasing levels of waste). The relationship between the relative threshold of interaction and the ``power of influence'' is not monotonic: the share of adults in bundles dominated by young consumers (and the variability of this outcome) increases with increasing open-mindedness of the latter.

Overall, the statistical sensitivity analysis shows that, considering each agent separately across 8,400 simulations, the absolute change in the individual waste level is negatively related to the initial waste level $W_0^i$ (agents wasting more initially tend to move less), $d_{int}^i$ (more open-minded agents move less, probably because they are subject to more contrasting influences), and $P_{intra}$ (more links inside the group result in smaller movements). Instead, the change is positively related to the initial opinion $O_0^i$ (as expected, since in our calibration opinions are more favourable to waste than behaviours), and the cognitive dissonance $d_{cd}^i$: agents willing to tolerate a larger mismatch between opinions and behaviours move more, which is partly counterintuitive and an interesting emerging dynamic of the model.

The stylised facts described in this subsection allow to draw useful recommendations. First, a small degree of mixing between socio-demographic groups (in this case young and adult consumers) is enough to activate reciprocal influence mechanisms. Virtuous (but potentially also wasteful) opinions spread more easily when ``points of contact'' are identified among reluctant people, and these are empowered of convincing their fellows, than in case of persistent and widespread campaigns. The benefits of a marginal increase in the degree of mixing beyond $P_{intra}=0.8$ (i.e., when young consumer have 20\% of their links with adults, and vice versa) are not worth the likely increasing costs of promoting interaction opportunities. Second, if two socio-demographic groups have different degrees of open-mindedness, the outcome in terms of waste levels can be non-trivial, as less wasteful individuals can also be attracted towards less virtuous behaviours. While a certain open-mindedness of wasteful people is a necessary condition to achieve a reduction in waste levels, too high willingness to interact can have a counterproductive effect. At the limit, if two groups present the same interaction threshold, the average levels of waste in the population do not change significantly as a result of their encounter. Therefore, policymakers should target their campaigns carefully to avoid recoil effects.

\subsection{The role of committed individuals}
In real-world situations, populations made up of clear-cut groups are rarely observed. Rather, a homogeneous population may include small groups of individuals who campaign for specific issues, such as food waste reduction. Given the specific dynamics involving extreme opinions highlighted in the literature (\citealt{deffuant2002can}), it is crucial to explore the power of committed individuals in convincing their peers to converge toward their behaviour.

We now generate populations that, in line with the parameters in Table \ref{committedParameters}, include 5.4\% of committed agents. These take an ``extreme'' stance on food waste, which translates into low initial food waste actions and opinions. As described in the calibration section, the values of their $W_0^i$ and $O_0^i$ are extracted from uniform distributions centred around $\overline{W_0^{C}}$ and $\overline{O_0^{C}}$ respectively, and with half-range $\sigma^{C}\in[0;0.2]$. Committed agents are also assumed to barely change their opinion, and only when engaging with individuals whose opinions are close to theirs. This translates into a small interaction threshold ($0<d_{int}^C<0.05$). In turn, the initial food waste levels of ordinary agents are calibrated using the procedure described in the Model Calibration subsection, and their open-mindedness is set at $d_{int}^{Ord}= 0.1$. Again, for each agent the threshold of ``cognitive dissonance'' is equal to \(\lvert W_i^0-O_i^0\rvert\), and thus smaller for committed agents by construction. Finally, the network is supposed to be fully mixed ($P_{intra}=0.5$), meaning that each agent has an equal probability to be connected with agents from either group.

\begin{figure}[h]
\centering
\includegraphics[width=1\linewidth]{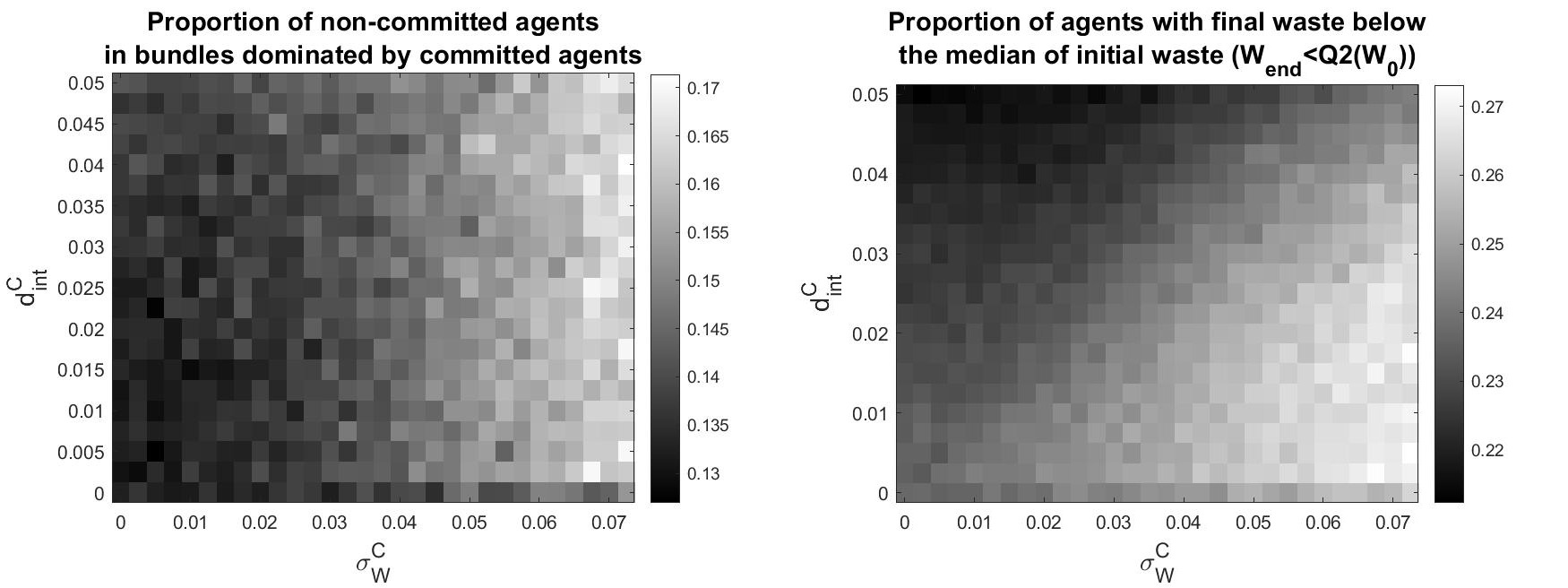}
\caption{\textbf{Committed vs ordinary consumers.} Left panel: Proportion of ordinary agents ending up in bundles dominated (see definition in the text) by committed agents. Right panel: Proportion of agents with $W_{end} < median(W_0)$. \textit{Notes}: The calibration parameters are set in line with Table \ref{committedParameters}. The colours of each cell are based on averages across 100 simulations for all possible combinations of $d_{int}^C\in[0;0.05]$ (in steps of 0.0025) and $\sigma^{C}\in[0;0.2]$ (in steps of 0.0025).}
\label{Fig3}
\end{figure}

To test the power of attraction of committed individuals, we assess the proportion of ordinary consumers ending up in bundles dominated by the former and the overall proportion of agents ending up with waste levels below the initial median of the population, both at $t_{max}$. We implement a statistical analysis by progressively changing the interaction threshold of committed individuals $d_{int}^C$ from 0 to 0.05 in steps of 0.0025, and the half-range of their initial uniformly distributed food waste opinions and behaviours ($\sigma^{C}$) from 0 to 0.2 in steps of 0.0025.\footnote{Since $\overline{W_0^{C}} < 0.2$ for all population, the actual range of $\sigma^{C}$ is smaller than 0.2.} The results for each possible combination of the two parameters are illustrated in Fig. \ref{Fig3}, while a statistical analysis is provided as Supplementary Material.

\begin{figure}[H]
\centering
\includegraphics[width=1\linewidth]{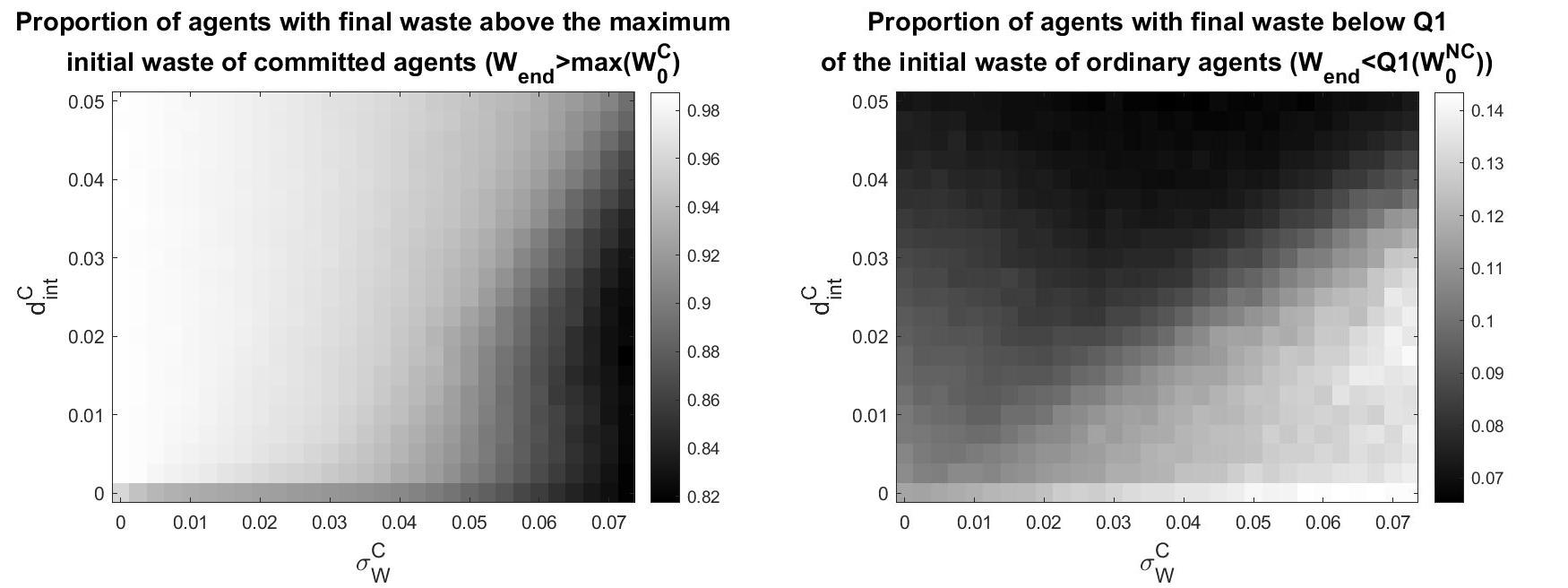}
\caption{\textbf{Additional analyses on the results of Fig. \ref{Fig3}.} Left panel: Proportion of agents with $W_{end} > max(W_0^{C})$. Right panel: Proportion of agents with $W_{end} < Q_1(W_0^{Ord})$.}
\label{FigS1}
\end{figure}

Like in the previous case study, to align with the calibration dataset, the opinions are initialised as more ``favourable'' to food waste compared to the behaviours. Furthermore, since the ``range of dishonesty'' that committed individuals are willing to accept is narrower on average, their behaviours are more likely to align with their opinions. Given these premises, the final waste levels are much higher than the initial ones: the proportion of agents ending up with $W_{end} < median(W_0)$ ranges between 22--27\% depending on the setting (right panel of Fig. \ref{Fig3}). This is still a better outcome than in the previous case study, where between 6--16\% of the agents ended up below $median(W_0)$ (upper-left panel of Fig. \ref{Fig2}), pointing to the beneficial impact of committed groups when specific conditions -- detailed below -- are verified. Another key stylised fact emerging from Fig. \ref{Fig3} is that the combination of open-mindedness and variability of opinions matters more than the value of these parameters separately. The only exception is when $d_{int}^{C}=0$ (i.e., when committed agents do not engage with anyone, illustrated in the bottom line of the two panels): in this case the value of $\sigma^{C}$ does not matter. Committed people are thus doomed to have no relevance beyond their own group if they are not willing to compromise with the rest of the population. The three-way relationship between the openness of mind $d_{int}^{C}$, the variability of food waste $\sigma^{C}$, and final waste $W_{end}$ is not linear. The ``best'' outcomes are achieved when the openness of mind is relatively low ($d_{int}^{C}<0.03$), and the variability of food waste for committed individuals relatively high ($\sigma^{C}>0.04$). When committed agents are more spread, they are more likely to attract ordinary agents located within their interaction threshold towards their positions; however, if their ``openness of mind'' is too large, they may be attracted towards higher food waste levels in turn.\footnote{The results in Fig. \ref{FigS1} confirm these finding, being either symmetrical (left panel) or analogous (right panel) to the right panel of Fig. \ref{Fig3}.} The statistical sensitivity analysis reported in Supplementary Material confirms this: beside the significant coefficients linking the proportion of agents with $W_{end} < median(W_0)$ to $d_{int}^{C}$ (negative) and $\sigma^{C}$ (positive), the interaction between the two parameters also yields a significant coefficient (negative). Compared to zero as a baseline, increasing values of $d_{int}^{C}$ yield first positive and then increasing negative results, while the impact of $\sigma^{C}$ is positive and increasing.

A similarly complex pattern is shown in the left panel of Fig. \ref{Fig3}, although $d_{int}^{C}$ matters less in this case. The proportion of ordinary agents ending up in bundles dominated by committed agents is larger when the variability of the latter's food waste is relatively high ($\sigma^{C}>0.04$), while for low variability relatively high levels of the openness of mind ($d_{int}^{C}>0.035$) generate slightly better outcomes. The statistical sensitivity analysis confirms these observations, as both parameters are significantly and positively related to the variable plotted in the left panel of Fig. \ref{Fig3}, but their interaction term is non-significant. Equally, any value of $d_{int}^{C}>0$ yields a positive power of attraction of committed agents compared to $d_{int}^{C}=0$, but with no clear pattern, while increasing $\sigma^{C}$ yields a clearly increasing power of attraction. It is worth remembering that a bundle is dominated by committed agents when their proportion is larger than their proportion in the overall population plus two standard deviations. This proportion is 5.4\%; hence, these bundles can include a large majority of ordinary agents but are still subject to the influence of committed individuals. Indeed, most settings with a large proportion of ordinary agents ending up in bundles dominated by their counterpart are also characterised by lower final levels of waste (right panel of Fig. \ref{Fig3}).

These results suggest that under certain conditions -- such as sufficient within-group variation $\sigma^{C}$ and positive but not-too-large openness of mind $d_{int}^{C}$ -- committed consumers can attract (or retain) a considerable share of individuals close to their ideas. We thus reproduce a result similar to \cite{deffuant2002can}, whose model adopts relative agreement between pairs of individuals to simulate the emerging and the influence of extremism, but translate and embed it to the problem of food waste. To maximize their effectiveness in spreading virtuous food waste behaviours, committed individuals should allow a certain variability of opinions within their group and, at the same time, be willing to engage with people whose opinions are different from theirs but not too much. This result aligns well with \cite{watts2007}, which found that the presence of critical masses of easily influenceable individuals is more important than the presence of influencers in generating a cascade effect. It is also worth mentioning that more variability among committed individuals implies slightly higher food waste, suggesting that there is a trade-off between power of persuasion and the achievable unitary improvements.

Being the first model to study the impact of social interactions on food waste generation though extensive simulations, there is a lack of studies for comparison. More in general, our results suggest that, in line with \cite{huh2014social}, social embeddedness and peers' choices can influence behaviour even when the latter shows a high heterogeneity and is private (i.e., takes place at home) like consumer food waste. In line with the literature on resource conservation, networks and similarity emerge as moderators between social influence and behavioural change (\citealt{abrahamse2013social}), whereas in line with the social psychology theory, also in the food waste domain habits exert a sticky influence, i.e. resist change (\citealt{verplanken1998habit}). Interaction opportunities among different socio-demographic groups should thus be considered by policies tackling food waste.

 \section{Conclusions and policy implications}\label{sec:conclusion} 
The present model, in its simplicity, has uncovered interesting stylized facts concerning the influence of social interactions on private environmental behaviours, providing insights into behaviour change critical for the design of effective policy interventions. To this end, we have explored the effects of social interactions on \textit{stated} opinions and \textit{private} food waste behaviours by means of an ABM. We built two numerical case studies of heterogeneous populations by calibrating the model with Italian data on household food waste. First, we assumed our populations to be made of adult and young consumers, with the former wasting less on average than the latter, and observed the impact on food waste of their mixing. Then, we assessed the convincing power of a small group of committed individuals. Indeed, in a population divided into fairly homogeneous groups, policy interventions may choose to target specific groups, and inter-group interactions are likely to scale up their impact. 

Our simulations show that even a low degree of mixing (i.e., a small number of inter-group interactions) is sufficient to convince individuals across groups, particularly those showing ``extreme'' behaviours to converge towards more ``central'' positions. 
Considering that the degree of mixing between adult and young consumers is higher where interactions between different generations are more frequent, these results suggest that extended households and houses of multiple occupancy could be characterized by a smaller number of individuals showing ``extreme'' food waste behaviours. While we calibrated our model using real food waste data, the network linking the agents is an Erd{\"o}s-R{\'e}nyi random network with density $\lambda=0.5$ -- a value which is not based on a real population. In Italy, the interactions between people of different age are more common than in other EU counties (\citealt{albertini2016ageing}), suggesting that the patterns of behaviour diffusion can be location-specific, and our findings do not necessarily extend to all countries when local network structures are factored in. 

Additionally, we found that a small number of committed consumers can effectively attract (or retain) other consumers close to their position if they show some within-group variability of opinion and are willing to consider different opinions but not too different. Open-mindedness alone does not allow them to reach out to a relevant number of individuals, while variability alone is not enough to ensure that they engage with ordinary people without being attracted towards them instead. The optimum in terms of convincing power is achieved when the opinions of committed individuals are spread across a certain range, and each of them engage with people located in their ``neighbourhood'' in terms of opinions. This allows to conclude that, when food waste is concerned, normative appeals are most effective when they describe group behaviours occurring in the immediate situational circumstances of target individuals, as shown for other topics by \cite{goldstein2008room}. Behavioural change is mediated by perceived similarity (\citealt{festinger1954theory}), and a certain degree of heterogeneity ensures that some committed individuals are closer to ordinary individuals. The level of polarisation and ``sectarianism'' (e.g., of radical pro-environmental groups) can differ between societies. While our committed agents were identified based on declared food-waste behaviours in Italy, other countries can show different group sizes and inter-group dynamics.

From the research point of view, this paper innovates with respect to previous works adopting a bounded confidence framework in random pairwise encounters (\citealt{weisbuch2002meet, deffuant2002can, goldstein2008room}) by including an ``interaction threshold'' -- agents with distant opinions are unlikely to continue to interact with, and be influenced by, each other. Future extensions might include a counterreaction, i.e. a movement in the opposite direction when others' opinions are too distant from one's prior. We have also introduced a novel ``cognitive dissonance'' threshold to account for the gap between \textit{stated} opinions and \textit{private} behaviours.

Our results allow to draw recommendations for the design of more effective policy tackling consumer food waste and associated economic and environmental costs. We recommend interventions that present food waste reduction and prevention as the social norm. Since many factors hinder people self-regulatory capacity when food is concerned, voluntary ``small step'' approaches based on nudges should complement policy strategies relying on regulation or market incentives (\citealt{roberto2014need}). The literature on recycling suggests that economic measures penalising private food waste and making it visible could incentivise the reduction of household waste over time. For instance, some Swiss municipalities that have been charging waste disposal costs per bag have seen a reduction of private waste by 40\% (\citealt{carattini2018taxing}). 

Based on our simulations, and as already found by \cite{pineda2015mass} for mass media, it is recommendable for interventions nudging consumers towards less wasteful behaviours to avoid being too insistent or frequent. Indeed, a limited number of contacts seems to be enough to induce change; in turn, too insistent contacts could activate people's ``confirmation bias'', or increase their tolerance towards ``cognitive dissonance'', as they would be forced to sustain in public an opinion they do not abide by in private (\citealt{vetter2016possibilities}).

Experimental studies found that communication campaigns targeting environmental behavioural change should communicate preferably clear, achievable and easy-to-achieve messages (\citealt{vetter2016possibilities}). Based on these and on our findings, we recommend communication campaigns to be driven by practical goals, and adapted to the priors of their (potentially different) target groups without giving the impression of pursuing uncompromising principles. This is also in line with \cite{goldstein2008room}, who found the use of descriptive norms more effective than traditional appeals focused solely on environmental protection. Communicating the potential for reducing household expenses thanks to better food management is expected to yield a positive impact (\citealt{priefer2016food, schanes2018food, zamri2020delivery}). Instead, targeting only ``extreme'' wasters or individuals who share similar (virtuous) opinions are ineffective strategies.

Favouring the opening up of groups of committed individuals to people located in their neighbourhood in terms of opinions is also key. Another type of intervention we recommend is thus to create enabling conditions for different groups -- age and environmental behaviour-wise -- to exchange opinions on food management, including food waste. Community centres, charitable and faith-based organisations are examples of contexts where people of various backgrounds and demographics meet and share experiences. Initiatives and events held at such organisations could trigger intergenerational (and intergroup) influence and favour the active redistribution and reuse of food that is likely to go to waste (\citealt{mousa2017organizations}). 

The spaces for encounters should be -- with priority -- physical spaces, due to the higher potential of face-to-face interactions to drive behavioural change. Events and training activities focused on better household food management could be organised in shopping centres, supermarkets, and local markets. While less food waste at home could result in reduced food purchases, retail companies could benefit from such initiatives through improved trust from their clients -- particularly small shops and markets, where personalised relationships are the norm -- and better reputation among environmentally-concerned consumers. School-based information and training focused on food waste are also likely to initiateboth intergenerational and intergroup influence, as students from different households interact, exchange opinions, and take the new ideas they are exposed to back to their households (\citealt{ballantyne2001program, vaughan2003effect, duvall2007review, damerell2013child, marchini2020can}).

As exemplified by the simulation case studies presented, the focus of our analysis was toward understanding the \textit{qualitative} directions of food waste production under different conditions concerning the structure of social interactions -- conditions that can be addressed through policy measures. For this reason, we did not aim to produce quantitative predictions, but rather study the sign of the changes. From the methodological point of view, ABMs are well suited to explore social systems building macro-outcomes from the bottom-up, and identifying how different micro-level assumptions influence the macro-outcomes (\citealt{schelling2006micromotives}). The food waste behaviour of individuals is arguably a complex system, whose determinants are many and only partially understood. Accordingly, we use our model to assess the direction and relative effectiveness of different variations to the status quo. While the structural dynamics uncovered remain valid, and could possibly be extended to other country settings, the results in terms of food waste levels should thus be considered carefully. Indeed, calibration was implemented using questionnaire data from Italy, and beside country specificities, the literature has shown that people tend to misestimate (usually underestimate) their food waste (\citealt{giordano2018questionnaires}). For the same reason, the values of the threshold of ``cognitive dissonance'' (gap between actions and opinions) could also be biased. These calibration issues, together with the requirement of data on socio-psychological constructs which are currently unavailable, represent main limitations of our model. Another limitation relates to the simplifying assumptions that we have made to make a complex phenomenon more tractable.

Rather than challenging the theoretical background underpinning the model, these caveats call for taking additional steps to better ground it into real-world situations. First, consumer food waste could be measured using more reliable methods, such as diaries or waste sorting analysis, and the resulting data used to simulate the impact of policy interventions. For example, one-to-many interactions like informational campaigns via TV or on online platforms could be simulated, as done in previous opinion dynamic studies (\citealt{pineda2015mass}. Second, competing motivations related to food could be introduced, e.g. a desire to achieve a status through food abundance (\citealt{piras2021community}). These motivations could interact through a utility function, creating trade-offs, and also be influenced by social network. Third, efforts are required to identify the structure of social interactions specific of the setting considered. Future studies could test the impact of alternative network structures, such as the Barab{\'a}si-Albert scale-free network topology (\citealt{barabasi1999emergence}), reproducing key features of human interaction networks in large-scale populations. Calibration could be implemented based on different dychotomies, e.g. between income earners and non-earners, or between social classes, given the role of income in consumer food waste generation (\citealt{SettietAl2016}). Finally, intra-household dynamics could be introduced by considering both intra-household networks and external networks. This would require information about food waste opinions of each member, while the food waste outcome would be the same for the entire household, possibly obtained through an averaging mechanism. Our ABM represents a valuable baseline where to build such additional dynamics.

\section*{Acknowledgements}\label{sec:acknowledge}
This research received funding from the Horizon 2020 Framework Programme of the European Union (project REFRESH ``Resource Efficient Food and dRink for the Entire Supply cHain'', Grant Agreement 641933) and the Scottish Government's Rural and Environment Science and Analytical Services Division (RESAS RD3.2.4 ``Food Culture and Dietary Choice''). The views reflected in this paper represent the professional views of the authors and do not necessarily reflect the views of the funders.

\section*{Appendix}\label{sec:appendix}

\begin{figure}[H]
\centering
\includegraphics[width=1\linewidth]{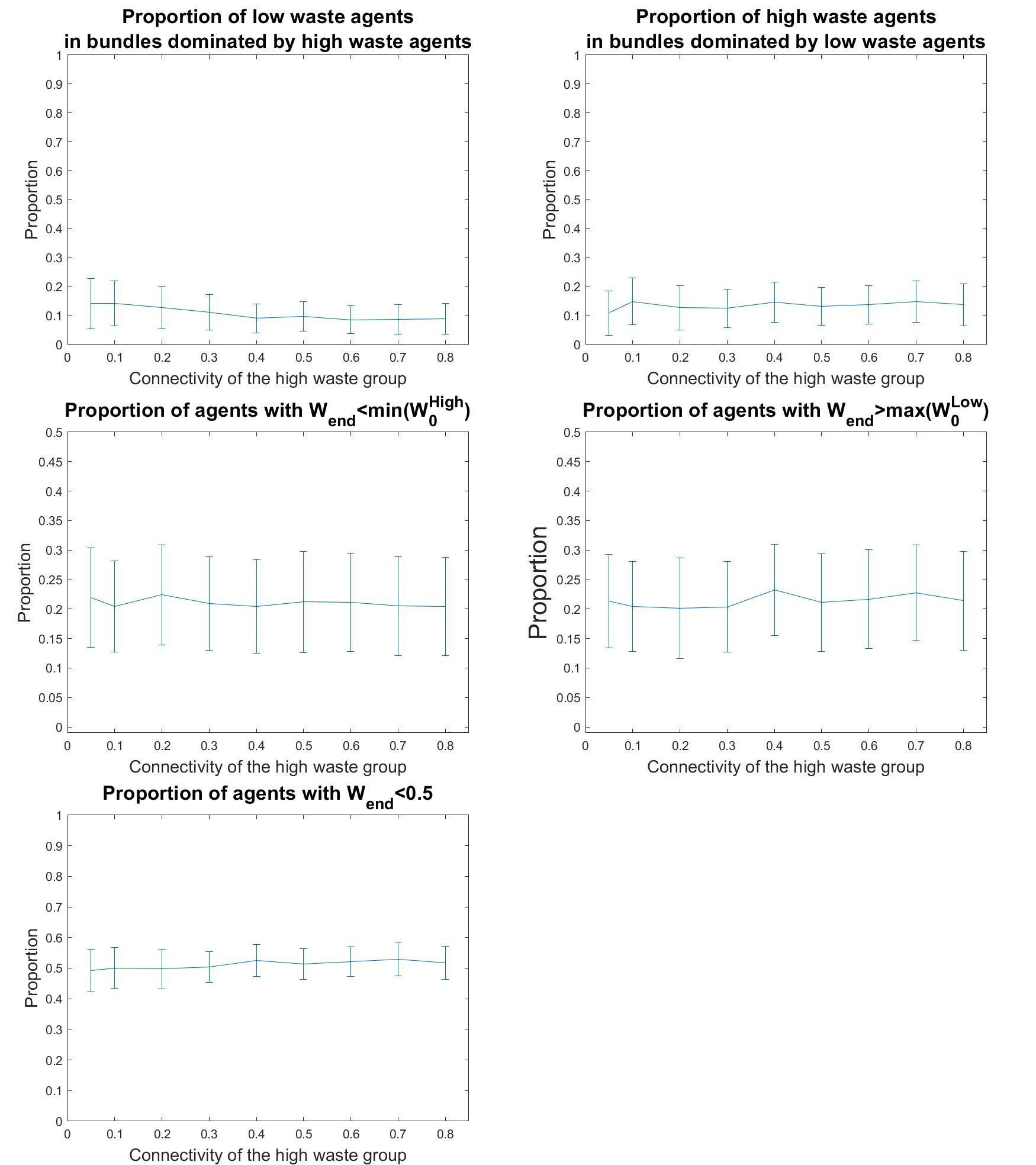}
\caption{\textbf{Impact of connectivity.} Robustness checks for the relative variation of the network degree $\lambda_j$ across groups. Consider two groups of equal sizes; the first group has a connectivity $\lambda=0.1$; the connectivity of the second group is varied. The first group's initial waste levels follow a triangular distribution $Tri[0.33 \pm 0.33]$, those of the second group a triangular distribution $Tri[0.67 \pm 0.33]$. This figure confirm the robustness of the results in the main text. }
\label{FigS2}
\end{figure}

\begin{figure}[H]
\centering
\includegraphics[width=1\linewidth]{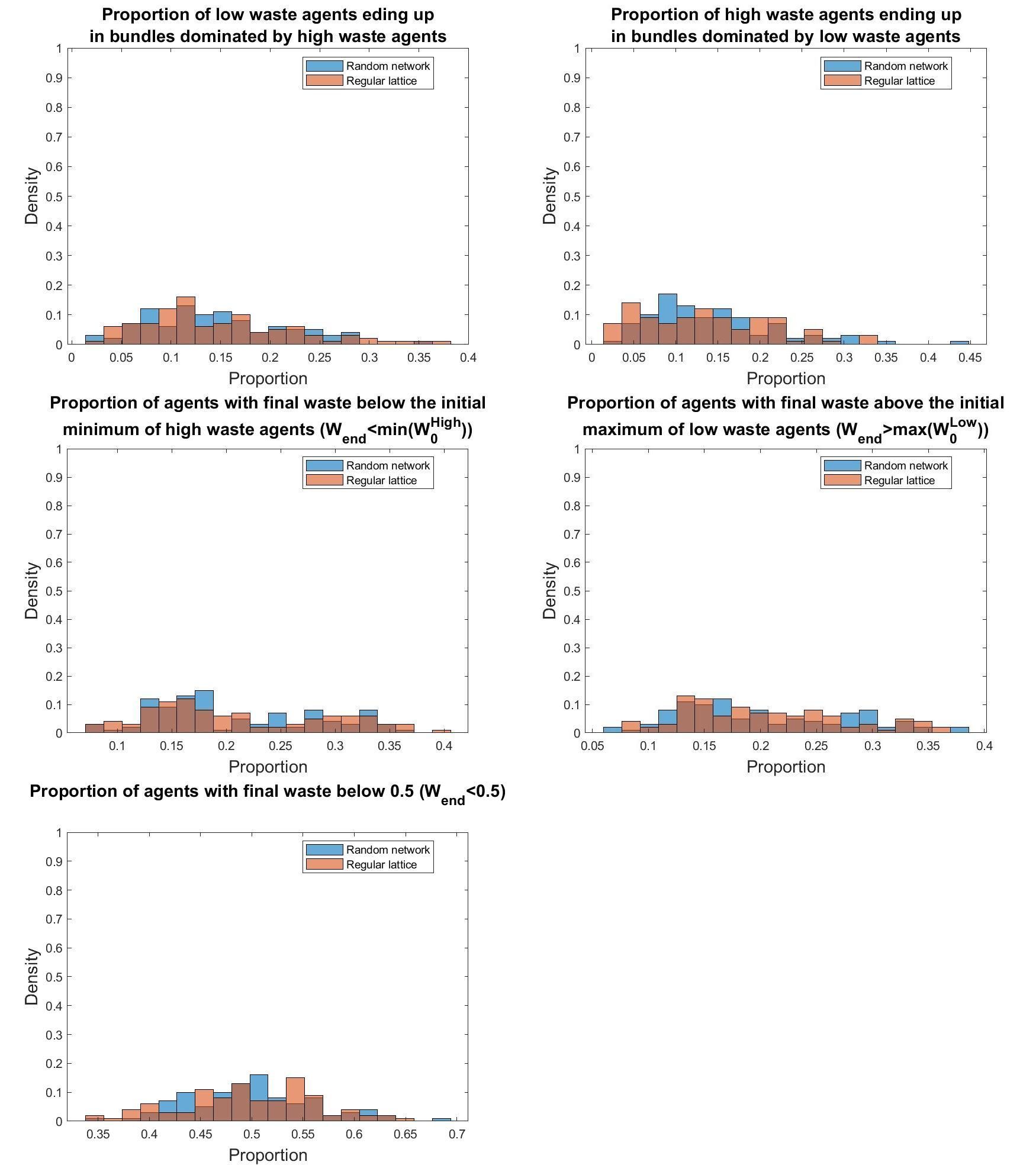}
\caption{\textbf{Lattices and Random Networks.} Consider two equally-sized groups with the same degree of mixing. The first group's initial waste levels follow a triangular distribution $Tri[0.33 \pm 0.33]$; those of the second group a triangular distribution $Tri[0.67 \pm 0.33]$. This is the only difference between the groups.}
\label{FigS3}
\end{figure}

\nolinenumbers

\bibliography{bibliography}{}

\begin{thebibliography}{101}
\providecommand{\natexlab}[1]{#1}
\providecommand{\url}[1]{\texttt{#1}}
\expandafter\ifx\csname urlstyle\endcsname\relax
  \providecommand{\doi}[1]{doi: #1}\else
  \providecommand{\doi}{doi: \begingroup \urlstyle{rm}\Url}\fi

\bibitem[Abelson et~al.(1968)Abelson, Aronson, McGuire, Newcomb, Rosenberg, and
  Tannenbaum]{abelson1968theories}
R.P. Abelson, E.E. Aronson, W.J. McGuire, T.M. Newcomb, M.J. Rosenberg, and
  P.H. Tannenbaum.
\newblock \emph{Theories of cognitive consistency: A sourcebook.}
\newblock Chicago, 1968.

\bibitem[Abrahamse and Steg(2013)]{abrahamse2013social}
W.~Abrahamse and L.~Steg.
\newblock Social influence approaches to encourage resource conservation: a
  meta-analysis.
\newblock \emph{Global environmental change}, 23\penalty0 (6):\penalty0
  1773--1785, 2013.

\bibitem[Ajzen(1991)]{ajzen1991theory}
Icek Ajzen.
\newblock The theory of planned behavior.
\newblock \emph{{O}rganizational {B}ehavior and {H}uman {D}ecision
  {P}rocesses}, 50\penalty0 (2):\penalty0 179--211, 1991.

\bibitem[Albertini(2016)]{albertini2016ageing}
Marco Albertini.
\newblock \emph{Ageing and family solidarity in Europe: patterns and driving
  factors of intergenerational support}.
\newblock The World Bank, Washington, DC, 2016.

\bibitem[Argo et~al.(2012)Argo, Dahl, and Manchanda]{argo2012}
J.J. Argo, D.W. Dahl, and R.V. Manchanda.
\newblock The influence of a mere social presence in a retail context.
\newblock \emph{{J}ournal of {C}onsumer {R}esearch}, 32\penalty0 (2):\penalty0
  207--212, 2012.

\bibitem[Ballantyne et~al.(2001)Ballantyne, Fien, and
  Packer]{ballantyne2001program}
Roy Ballantyne, John Fien, and Jan Packer.
\newblock Program effectiveness in facilitating intergenerational influence in
  environmental education: Lessons from the field.
\newblock \emph{The journal of environmental education}, 32\penalty0
  (4):\penalty0 8--15, 2001.

\bibitem[Barab{\'a}si and Albert(1999)]{barabasi1999emergence}
A.L. Barab{\'a}si and R~Albert.
\newblock Emergence of scaling in random networks.
\newblock \emph{{S}cience}, 286\penalty0 (5439):\penalty0 509--512, 1999.

\bibitem[Barr(2007)]{barr2007factors}
S~Barr.
\newblock Factors influencing environmental attitudes and behaviors: A {UK}
  case study of household waste management.
\newblock \emph{{E}nvironment and {B}ehavior}, 39\penalty0 (4):\penalty0
  435--473, 2007.

\bibitem[Bolton and Alba(2012)]{bolton2012less}
L.E. Bolton and J.W. Alba.
\newblock When less is more: {c}onsumer aversion to unused utility.
\newblock \emph{{J}ournal of {C}onsumer {P}sychology}, 22\penalty0
  (3):\penalty0 369--383, 2012.

\bibitem[Bowles and Gintis(2004)]{bowles2004evolution}
S~Bowles and H~Gintis.
\newblock The evolution of strong reciprocity: cooperation in heterogeneous
  populations.
\newblock \emph{Theoretical Population Biology}, 65\penalty0 (1):\penalty0
  17--28, 2004.

\bibitem[Bublitz et~al.(2010)Bublitz, Peracchio, and Block]{bublitz2010}
M.G. Bublitz, L.A. Peracchio, and L.G. Block.
\newblock Why did {I} eat that? {P}erspectives on food decision making and
  dietary restraint.
\newblock \emph{{J}ournal of {C}onsumer {P}sychology}, 20\penalty0
  (3):\penalty0 239--258, 2010.

\bibitem[Canali et~al.(2017)Canali, Amani, Aramyan, Gheoldus, Moates,
  {\"O}stergren, Silvennoinen, Waldron, and Vittuari]{canali2016food}
M~Canali, P~Amani, L~Aramyan, M~Gheoldus, G~Moates, K~{\"O}stergren,
  K~Silvennoinen, K~Waldron, and M~Vittuari.
\newblock Food waste drivers in {E}urope, from identification to possible
  interventions.
\newblock \emph{Sustainability}, 9\penalty0 (1):\penalty0 37, 2017.

\bibitem[Carattini et~al.(2018)Carattini, Baranzini, and
  Lalive]{carattini2018taxing}
Stefano Carattini, Andrea Baranzini, and Rafael Lalive.
\newblock Is taxing waste a waste of time? evidence from a supreme court
  decision.
\newblock \emph{Ecological Economics}, 148:\penalty0 131--151, 2018.

\bibitem[Carroni et~al.(2020)Carroni, Pin, and Righi]{CarroniPinRighi}
E~Carroni, P~Pin, and S~Righi.
\newblock Bring a friend! privately or publicly?
\newblock \emph{Management Science}, 66\penalty0 (5):\penalty0 2269--2290,
  2020.

\bibitem[Cecere et~al.(2014)Cecere, Mancinelli, and Mazzanti]{cecere2014waste}
G~Cecere, S~Mancinelli, and M~Mazzanti.
\newblock Waste prevention and social preferences: {T}he role of intrinsic and
  extrinsic motivations.
\newblock \emph{Ecological Economics}, 107:\penalty0 163--176, 2014.

\bibitem[Cialdini et~al.(1990)Cialdini, Reno, and Kallgren]{cialdini1990focus}
R.B. Cialdini, R.R. Reno, and C.A. Kallgren.
\newblock A focus theory of normative conduct: recycling the concept of norms
  to reduce littering in public places.
\newblock \emph{{J}ournal of {P}ersonality and {S}ocial {P}sychology},
  58\penalty0 (6):\penalty0 1015--1026, 1990.

\bibitem[Cornil et~al.(2014)Cornil, Ordabayeva, Kaiser, Weber, and
  Chandon]{corniletal2014acuity}
Y~Cornil, N~Ordabayeva, U~Kaiser, B~Weber, and P~Chandon.
\newblock The acuity of vice: {A}ttitude ambivalence improves visual
  sensitivity to increasing portion sizes.
\newblock \emph{{J}ournal of {C}onsumer {P}sychology}, 24\penalty0
  (2):\penalty0 177--187, 2014.

\bibitem[Damerell et~al.(2013)Damerell, Howe, and
  Milner-Gulland]{damerell2013child}
Peter Damerell, Caroline Howe, and Eleanor~J Milner-Gulland.
\newblock Child-orientated environmental education influences adult knowledge
  and household behaviour.
\newblock \emph{Environmental Research Letters}, 8\penalty0 (1):\penalty0
  015016, 2013.

\bibitem[Deffuant et~al.(2002)Deffuant, Amblard, Weisbuch, and
  Faure]{deffuant2002can}
G~Deffuant, F~Amblard, G~Weisbuch, and T~Faure.
\newblock How can extremism prevail? {A} study based on the relative agreement
  interaction model.
\newblock \emph{{J}ournal of {A}rtificial {S}ocieties and {S}ocial
  {S}imulation}, 5\penalty0 (4), 2002.

\bibitem[Dreber et~al.(2008)Dreber, Rand, Fudenberg, and
  Nowak]{dreber2008winners}
A~Dreber, D.G. Rand, D~Fudenberg, and M.A. Nowak.
\newblock Winners don't punish.
\newblock \emph{Nature}, 452\penalty0 (7185):\penalty0 348--351, 2008.

\bibitem[Duvall and Zint(2007)]{duvall2007review}
Jason Duvall and Michaela Zint.
\newblock A review of research on the effectiveness of environmental education
  in promoting intergenerational learning.
\newblock \emph{The Journal of Environmental Education}, 38\penalty0
  (4):\penalty0 14--24, 2007.

\bibitem[Erd{\"o}s and R{\'e}nyi(1959)]{erdos1959random}
P~Erd{\"o}s and A~R{\'e}nyi.
\newblock On random graphs, {I}.
\newblock \emph{{P}ublicationes {M}athematicae ({D}ebrecen)}, 6:\penalty0
  290--297, 1959.

\bibitem[Evans(2011)]{evans2011blaming}
D~Evans.
\newblock Blaming the consumer--once again: {T}he social and material contexts
  of everyday food waste practices in some {E}nglish households.
\newblock \emph{Critical Public Health}, 21\penalty0 (4):\penalty0 429--440,
  2011.

\bibitem[Fehr and G{\"a}chter(2005)]{fehr2005human}
E~Fehr and S~G{\"a}chter.
\newblock Human behaviour: Egalitarian motive and altruistic punishment
  (reply).
\newblock \emph{Nature}, 433\penalty0 (7021):\penalty0 E1--E2, 2005.

\bibitem[Festinger(1954)]{festinger1954theory}
L~Festinger.
\newblock A theory of social comparison processes.
\newblock \emph{Human relations}, 7\penalty0 (2):\penalty0 117--140, 1954.

\bibitem[Festinger(1962)]{festinger1962}
Leon Festinger.
\newblock Cognitive dissonance.
\newblock \emph{{S}cientific {A}merican}, 207\penalty0 (4):\penalty0 93--107,
  1962.

\bibitem[Fowler(2005)]{fowler2005altruistic}
J.H. Fowler.
\newblock Altruistic punishment and the origin of cooperation.
\newblock \emph{Proceedings of the National Academy of Sciences of the United
  States of America}, 102\penalty0 (19):\penalty0 7047--7049, 2005.

\bibitem[Fowler et~al.(2005)Fowler, Johnson, and
  Smirnov]{fowler2005egalitarian}
J.H. Fowler, T~Johnson, and O~Smirnov.
\newblock Egalitarian motive and altruistic punishment.
\newblock \emph{Nature}, 433\penalty0 (10.1038):\penalty0 137--140, 2005.

\bibitem[Gaiani et~al.(2018)Gaiani, Caldeira, Adorno, Segr{\`e}, and
  Vittuari]{gaiani2018food}
Silvia Gaiani, Sandra Caldeira, Valentina Adorno, Andrea Segr{\`e}, and Matteo
  Vittuari.
\newblock Food wasters: Profiling consumers’ attitude to waste food in italy.
\newblock \emph{Waste Management}, 72:\penalty0 17--24, 2018.

\bibitem[Galam(2008)]{galam2008sociophysics}
S~Galam.
\newblock Sociophysics: a review of {G}alam models.
\newblock \emph{International Journal of Modern Physics C}, 19\penalty0
  (03):\penalty0 409--440, 2008.

\bibitem[Galeotti(2010)]{galeotti2010talking}
A~Galeotti.
\newblock Talking, searching, and pricing.
\newblock \emph{{I}nternational {E}conomic {R}eview}, 51\penalty0 (4):\penalty0
  1159--1174, 2010.

\bibitem[Giordano et~al.(2018)Giordano, Piras, Boschini, and
  Falasconi]{giordano2018questionnaires}
C~Giordano, S~Piras, M~Boschini, and L~Falasconi.
\newblock Are questionnaires a reliable method to measure food waste? a pilot
  study on italian households.
\newblock \emph{British Food Journal}, 2018.

\bibitem[Goldstein et~al.(2008)Goldstein, Cialdini, and
  Griskevicius]{goldstein2008room}
N.J. Goldstein, R.B. Cialdini, and V.~Griskevicius.
\newblock A room with a viewpoint: {U}sing social norms to motivate
  environmental conservation in hotels.
\newblock \emph{{J}ournal of {C}onsumer {R}esearch}, 35\penalty0 (3):\penalty0
  472--482, 2008.

\bibitem[Graham-Rowe and Sparks(2015)]{graham2015predicting}
D.C. Graham-Rowe, E.and~Jessop and P.~Sparks.
\newblock Predicting household food waste reduction using an extended theory of
  planned behaviour.
\newblock \emph{Resources, Conservation and Recycling}, 101:\penalty0 194--202,
  2015.

\bibitem[Graham-Rowe et~al.(2014)Graham-Rowe, Jessop, and
  Sparks]{graham2014identifying}
E~Graham-Rowe, D.C. Jessop, and P~Sparks.
\newblock Identifying motivations and barriers to minimising household food
  waste.
\newblock \emph{Resources, Conservation and Recycling}, 84:\penalty0 15--23,
  2014.

\bibitem[Grainger et~al.(2018{\natexlab{a}})Grainger, Aramyan, Logatcheva,
  Piras, Righi, Setti, Vittuari, and Stewart]{grainger2018use}
Matthew~James Grainger, Lusine Aramyan, Katja Logatcheva, Simone Piras, Simone
  Righi, Marco Setti, Matteo Vittuari, and Gavin~Bruce Stewart.
\newblock The use of systems models to identify food waste drivers.
\newblock \emph{Global food security}, 16:\penalty0 1--8, 2018{\natexlab{a}}.

\bibitem[Grainger et~al.(2018{\natexlab{b}})Grainger, Aramyan, Piras, Quested,
  Righi, Setti, Vittuari, and Stewart]{grainger2018model}
Matthew~James Grainger, Lusine Aramyan, Simone Piras, Thomas~Edward Quested,
  Simone Righi, Marco Setti, Matteo Vittuari, and Gavin~Bruce Stewart.
\newblock Model selection and averaging in the assessment of the drivers of
  household food waste to reduce the probability of false positives.
\newblock \emph{PloS one}, 13\penalty0 (2):\penalty0 e0192075,
  2018{\natexlab{b}}.

\bibitem[Grainger and Stewart(2017)]{MattGavinAswer2017}
M.J. Grainger and G.B. Stewart.
\newblock The jury is still out on social media as a tool for reducing food
  waste a response to {Y}oung et al. (2017).
\newblock \emph{Resources, {C}onservation and {R}ecycling}, 122:\penalty0
  407--410, 2017.

\bibitem[Gramsci(1975)]{gramsci1975quaderni}
Antonio Gramsci.
\newblock Quaderni dal carcere.
\newblock \emph{Torino, Einaudi}, 1975.

\bibitem[Gunders(2012)]{gunders2012wasted}
D~Gunders.
\newblock Wasted: How {A}merica is losing up to 40 percent of its food from
  farm to fork to landfill.
\newblock Technical report, Natural Resources Defense Council, 2012.

\bibitem[Gustavsson et~al.(2011)Gustavsson, Cederberg, Sonesson, Van~Otterdijk,
  and Meybeck]{gustavsson2011global}
J~Gustavsson, C~Cederberg, U~Sonesson, R~Van~Otterdijk, and A~Meybeck.
\newblock Global food losses and food waste.
\newblock Technical report, Food and Agriculture Organization of the United
  Nations, Roma, 2011.

\bibitem[Hegselmann et~al.(2002)Hegselmann, Krause,
  et~al.]{hegselmann2002opinion}
R~Hegselmann, U~Krause, et~al.
\newblock Opinion dynamics and bounded confidence models, analysis, and
  simulation.
\newblock \emph{Journal of Artificial Societies and Social Simulation},
  5\penalty0 (3), 2002.

\bibitem[Huh et~al.(2014)Huh, Vosgerau, and Morewedge]{huh2014social}
Y.E. Huh, J~Vosgerau, and C.K. Morewedge.
\newblock Social defaults: {O}bserved choices become choice defaults.
\newblock \emph{{J}ournal of {C}onsumer {R}esearch}, 41\penalty0 (3):\penalty0
  746--760, 2014.

\bibitem[Jackson(2005)]{jackson2005economics}
M.O. Jackson.
\newblock The economics of social networks.
\newblock In \emph{Social Science Working Paper 1237}. California {I}nstitute
  of {T}echnology, 2005.

\bibitem[Jackson(2010)]{jackson2010social}
M.O. Jackson.
\newblock \emph{Social and economic networks}.
\newblock Princeton {U}niversity {P}ress, 2010.

\bibitem[Jackson et~al.(2017)Jackson, Rogers, and Zenou]{Jackson2016}
M.O. Jackson, B.W. Rogers, and Y.~Zenou.
\newblock The economic consequences of social network structure.
\newblock Forthcoming, 2017.

\bibitem[Ji and Wood(2007)]{jiandwood2007}
M.F. Ji and W~Wood.
\newblock Purchase and consumption habits: Not necessarily what you intend.
\newblock \emph{{J}ournal of {C}onsumer {P}sychology}, 17\penalty0
  (1):\penalty0 261--276, 2007.

\bibitem[Kerr and Levine(2008)]{kerr2008detection}
N.L. Kerr and J.M. Levine.
\newblock The detection of social exclusion: Evolution and beyond.
\newblock \emph{Group Dynamics: Theory, Research, and Practice}, 12\penalty0
  (1):\penalty0 39, 2008.

\bibitem[Koivupuro et~al.(2012)Koivupuro, Hartikainen, Silvennoinen,
  Katajajuuri, Heikintalo, Reinikainen, and Jalkanen]{koivupuro2012influence}
H.K. Koivupuro, H.~Hartikainen, K.~Silvennoinen, J.M. Katajajuuri,
  N.~Heikintalo, A.~Reinikainen, and L.~Jalkanen.
\newblock Influence of socio-demographical, behavioural and attitudinal factors
  on the amount of avoidable food waste generated in finnish households.
\newblock \emph{{I}nternational {J}ournal of {C}onsumer {S}tudies}, 36\penalty0
  (2):\penalty0 183--191, 2012.

\bibitem[Kurzban and Leary(2001)]{kurzban2001evolutionary}
R.~Kurzban and M.R. Leary.
\newblock Evolutionary origins of stigmatization: {T}he functions of social
  exclusion.
\newblock \emph{Psychological Bulletin}, 127\penalty0 (2):\penalty0 187, 2001.

\bibitem[Lamberton et~al.(2013)Lamberton, Naylor, and
  Haws]{lamberton2013destination}
C.P. Lamberton, R.W. Naylor, and K.L. Haws.
\newblock Same destination, different paths: {W}hen and how does observing
  others' choices and reasoning alter confidence in our own choices?
\newblock \emph{{J}ournal of {C}onsumer {P}sychology}, 23\penalty0
  (1):\penalty0 74--89, 2013.

\bibitem[Lapinski and Rimal(2005)]{lapinski2005explication}
M.K. Lapinski and R.N. Rimal.
\newblock An explication of social norms.
\newblock \emph{Communication Theory}, 15\penalty0 (2):\penalty0 127--147,
  2005.

\bibitem[Lewin(1951)]{lewin1951field}
K~Lewin.
\newblock \emph{Field theory in social science}.
\newblock Harper, 1951.

\bibitem[Longhi(2013)]{longhi2013individual}
Simonetta Longhi.
\newblock Individual pro-environmental behaviour in the household context.
\newblock Technical report, ISER Working Paper Series, 2013.

\bibitem[Lorenz(2007)]{lorenz2007continuous}
J~Lorenz.
\newblock Continuous opinion dynamics under bounded confidence: A survey.
\newblock \emph{International Journal of Modern Physics C}, 18\penalty0
  (12):\penalty0 1819--1838, 2007.

\bibitem[Marchini and Macdonald(2020)]{marchini2020can}
Silvio Marchini and David~W Macdonald.
\newblock Can school children influence adults’ behavior toward jaguars?
  evidence of intergenerational learning in education for conservation.
\newblock \emph{Ambio}, 49\penalty0 (4):\penalty0 912--925, 2020.

\bibitem[McFerran et~al.(2010)McFerran, Dahl, Fitzsimons, and
  Morales]{mcferran2010}
B~McFerran, D.W. Dahl, G.J. Fitzsimons, and A.C. Morales.
\newblock I'll have what she's having: {E}ffects of social influence and body
  type on the food choices of others.
\newblock \emph{{J}ournal of {C}onsumer {R}esearch}, 36\penalty0 (6):\penalty0
  915--926, 2010.

\bibitem[Moller et~al.(2014a)Moller, Hanssen, Svanes, Hartikainen,
  Silvennoinen, Gustavsson, {\"O}stergren, Schneider, Soethoudt, Canali,
  et~al.]{moller2014standard}
H~Moller, O.J. Hanssen, E~Svanes, H~Hartikainen, K~Silvennoinen, J~Gustavsson,
  K~{\"O}stergren, F~Schneider, H~Soethoudt, M~Canali, et~al.
\newblock Standard approach on quantitative techniques to be used to estimate
  food waste levels.
\newblock Technical report, FUSIONS project, 2014a.
\newblock ISBN : 82-7520-723-1 978-82-7520-723-2.

\bibitem[M{\o}ller et~al.(2014b)M{\o}ller, Hanssen, Gustavsson, {\"O}stergren,
  Stenmarck, and Dekhtyar]{moller2014report}
H~M{\o}ller, O.J. Hanssen, J~Gustavsson, K~{\"O}stergren, A.A. Stenmarck, and
  P~Dekhtyar.
\newblock Report on review of (food) waste reporting methodology and practice.
\newblock Technical report, FUSIONS project, 2014b.
\newblock ISBN: 82-7520-713-4 978-82-7520-713-3.

\bibitem[Mousa and Freeland-Graves(2017)]{mousa2017organizations}
Tamara~Y Mousa and Jeanne~H Freeland-Graves.
\newblock Organizations of food redistribution and rescue.
\newblock \emph{Public health}, 152:\penalty0 117--122, 2017.

\bibitem[Mukhopadhyay et~al.(2008)Mukhopadhyay, Sengupta, and
  Ramanathan]{mukhopadhyay2008}
A~Mukhopadhyay, J~Sengupta, and S~Ramanathan.
\newblock Recalling past temptations: {A}n information-processing perspective
  on the dynamics of self-control.
\newblock \emph{{J}ournal of {C}onsumer {R}esearch}, 35\penalty0 (4):\penalty0
  586--599, 2008.

\bibitem[Nishi and Masuda(2013)]{NishiMasuda2013}
R~Nishi and N~Masuda.
\newblock Collective opinion formation model under {B}ayesian updating and
  confirmation bias.
\newblock \emph{{P}hysical {R}eview {E}}, 87:\penalty0 062123, Jun 2013.

\bibitem[Parizeau et~al.(2015)Parizeau, von Massow, and
  Martin]{parizeau2015household}
K~Parizeau, M~von Massow, and R~Martin.
\newblock Household--level dynamics of food waste production and related
  beliefs, attitudes, and behaviours in {G}uelph, {O}ntario.
\newblock \emph{Waste Management}, 35:\penalty0 207--217, 2015.

\bibitem[Pineda and Buend{\'\i}a(2015)]{pineda2015mass}
Mercedes Pineda and GM~Buend{\'\i}a.
\newblock Mass media and heterogeneous bounds of confidence in continuous
  opinion dynamics.
\newblock \emph{Physica A: Statistical Mechanics and its Applications},
  420:\penalty0 73--84, 2015.

\bibitem[Piras et~al.(2021)Piras, Pancotto, Righi, Vittuari, and
  Setti]{piras2021community}
Simone Piras, Francesca Pancotto, Simone Righi, Matteo Vittuari, and Marco
  Setti.
\newblock Community social capital and status: The social dilemma of food
  waste.
\newblock \emph{Ecological Economics}, 183:\penalty0 106954, 2021.

\bibitem[Priefer et~al.(2016)Priefer, J{\"o}rissen, and
  Br{\"a}utigam]{priefer2016food}
Carmen Priefer, Juliane J{\"o}rissen, and Klaus-Rainer Br{\"a}utigam.
\newblock Food waste prevention in europe--a cause-driven approach to identify
  the most relevant leverage points for action.
\newblock \emph{Resources, Conservation and Recycling}, 109:\penalty0 155--165,
  2016.

\bibitem[Putnam et~al.(1993)Putnam, Leonardi, and Nanetti]{putnam1993making}
Robert~D Putnam, Robert Leonardi, and Raffaella~Y. Nanetti.
\newblock \emph{Making democracy work: Civic traditions in modern Italy}.
\newblock Princeton University Press, Princeton, NJ, 1993.

\bibitem[Quested et~al.(2013)Quested, Marsh, Stunell, and
  Parry]{quested2013spaghetti}
T.E. Quested, E~Marsh, D~Stunell, and A.D. Parry.
\newblock Spaghetti soup: The complex world of food waste behaviours.
\newblock \emph{{R}esources, {C}onservation and {R}ecycling}, 79:\penalty0
  43--51, 2013.

\bibitem[Ravandi and Jovanovic(2019)]{ravandi2019impact}
Babak Ravandi and Nina Jovanovic.
\newblock Impact of plate size on food waste: Agent-based simulation of food
  consumption.
\newblock \emph{Resources, Conservation and Recycling}, 149:\penalty0 550--565,
  2019.

\bibitem[Roberto et~al.(2014)Roberto, Pomeranz, and Fisher]{roberto2014need}
C.A. Roberto, J.L. Pomeranz, and J.O. Fisher.
\newblock The need for public policies to promote healthier food consumption:
  {A} comment on {W}ansink and {C}handon (2014).
\newblock \emph{{J}ournal of {C}onsumer {P}sychology}, 24\penalty0
  (3):\penalty0 438--445, 2014.

\bibitem[Schanes et~al.(2018)Schanes, Dobernig, and G{\"o}zet]{schanes2018food}
Karin Schanes, Karin Dobernig, and Burcu G{\"o}zet.
\newblock Food waste matters-a systematic review of household food waste
  practices and their policy implications.
\newblock \emph{Journal of Cleaner Production}, 182:\penalty0 978--991, 2018.

\bibitem[Schelling(2006)]{schelling2006micromotives}
Thomas~C Schelling.
\newblock \emph{Micromotives and macrobehavior}.
\newblock WW Norton \& Company, 2006.

\bibitem[Schultz et~al.(2008)Schultz, Khazian, and Zaleski]{schultz2008using}
W.P. Schultz, A.M. Khazian, and A.C. Zaleski.
\newblock Using normative social influence to promote conservation among hotel
  guests.
\newblock \emph{Social {I}nfluence}, 3\penalty0 (1):\penalty0 4--23, 2008.

\bibitem[Secondi et~al.(2015)Secondi, Principato, and
  Laureti]{secondi2015household}
Luca Secondi, Ludovica Principato, and Tiziana Laureti.
\newblock Household food waste behaviour in {EU-27} countries: {A} multilevel
  analysis.
\newblock \emph{Food Policy}, 56:\penalty0 25--40, 2015.

\bibitem[Seebauer and Wolf(2017)]{seebauer2017disentangling}
Sebastian Seebauer and Angelika Wolf.
\newblock Disentangling household and individual actors in explaining private
  electricity consumption.
\newblock \emph{Energy Efficiency}, 10\penalty0 (1):\penalty0 1--20, 2017.

\bibitem[Sengupta et~al.(2002)Sengupta, Dahl, and Gorn]{sengupta2002misrep}
J~Sengupta, D.W. Dahl, and G.J. Gorn.
\newblock Misrepresentation in the consumer context.
\newblock \emph{{J}ournal of {C}onsumer {P}sychology}, 12\penalty0
  (2):\penalty0 69--79, 2002.

\bibitem[Setti et~al.(2016)Setti, Falasconi, Segr{\`e}, Cusano, and
  Vittuari]{SettietAl2016}
M.~Setti, L.~Falasconi, A.~Segr{\`e}, I.~Cusano, and M.~Vittuari.
\newblock Italian consumers' income and food waste behavior.
\newblock \emph{British Food Journal}, 118\penalty0 (7):\penalty0 1731--1746,
  2016.

\bibitem[Simmel(1908)]{simmel1908sociology}
G.~Simmel.
\newblock \emph{Sociology: investigations on the forms of sociation}.
\newblock Duncker \& Humblot, Berlin Germany, 1908.

\bibitem[Simpson et~al.(2012)Simpson, Griskevicius, and
  Rothman]{simpson2012relat}
J.A. Simpson, V~Griskevicius, and A.J. Rothman.
\newblock Consumer decisions in relationships.
\newblock \emph{{J}ournal of {C}onsumer {P}sychology}, 22\penalty0
  (3):\penalty0 304--314, 2012.

\bibitem[Sonesson et~al.(2005)Sonesson, Anteson, Davis, and
  Sj{\"o}d{\'e}n]{sonesson2005home}
U~Sonesson, F~Anteson, J~Davis, and P.O. Sj{\"o}d{\'e}n.
\newblock Home transport and wastage: {E}nvironmentally relevant household
  activities in the life cycle of food.
\newblock \emph{{AMBIO}: A {J}ournal of the {H}uman {E}nvironment}, 34\penalty0
  (4):\penalty0 371--375, 2005.

\bibitem[Stancu et~al.(2016)Stancu, Haugaard, and
  L{\"a}hteenm{\"a}ki]{stancu2016determinants}
V~Stancu, P~Haugaard, and L~L{\"a}hteenm{\"a}ki.
\newblock Determinants of consumer food waste behaviour: Two routes to food
  waste.
\newblock \emph{Appetite}, 96:\penalty0 7--17, 2016.

\bibitem[Stefan et~al.(2013)Stefan, van Herpen, Tudoran, and
  L{\"a}hteenm{\"a}ki]{stefan2013avoiding}
V~Stefan, E~van Herpen, A.A. Tudoran, and L~L{\"a}hteenm{\"a}ki.
\newblock Avoiding food waste by {R}omanian consumers: {T}he importance of
  planning and shopping routines.
\newblock \emph{{F}ood {Q}uality and {P}reference}, 28\penalty0 (1):\penalty0
  375--381, 2013.

\bibitem[Stenmarck et~al.(2016)Stenmarck, Jensen, Quested, Moates, Buksti,
  Cseh, Juul, Parry, Politano, Redlingshofer, et~al.]{stenmarck2011initiatives}
A~Stenmarck, C~Jensen, T~Quested, G~Moates, M~Buksti, B~Cseh, S~Juul, A~Parry,
  A~Politano, B~Redlingshofer, et~al.
\newblock Estimates of european food waste levels.
\newblock Technical report, FUSIONS project, 2016.
\newblock ISBN: 978-91-88319-01-2.

\bibitem[Stern(2000)]{stern2000new}
P.C. Stern.
\newblock New environmental theories: toward a coherent theory of
  environmentally significant behavior.
\newblock \emph{Journal of Social Issues}, 56\penalty0 (3):\penalty0 407--424,
  2000.

\bibitem[Tajfel and Turner(2004)]{tajfel2004social}
H~Tajfel and J.C. Turner.
\newblock \emph{The Social Identity Theory of Intergroup Behavior.}
\newblock Psychology Press, 2004.

\bibitem[Van~Baaren et~al.(2004)Van~Baaren, Holland, Kawakami, and
  Van~Knippenberg]{van2004mimicry}
R.B. Van~Baaren, R.W. Holland, K~Kawakami, and Ad~Van~Knippenberg.
\newblock Mimicry and prosocial behavior.
\newblock \emph{{P}sychological {S}cience}, 15\penalty0 (1):\penalty0 71--74,
  2004.

\bibitem[van Geffen et~al.(2016)van Geffen, van Herpen, and van
  Trijp]{Vangeffen2016standard}
L.E.J. van Geffen, E.~van Herpen, and J.C.M. van Trijp.
\newblock Causes \& determinants of consumers food waste. project report.
\newblock Technical report, EU Horizon 2020 REFRESH, 2016.
\newblock ISBN: 82-7520-713-4 978-82-7520-713-3.

\bibitem[Vaughan et~al.(2003)Vaughan, Gack, Solorazano, and
  Ray]{vaughan2003effect}
Christopher Vaughan, Julie Gack, Humberto Solorazano, and Robert Ray.
\newblock The effect of environmental education on schoolchildren, their
  parents, and community members: A study of intergenerational and
  intercommunity learning.
\newblock \emph{The Journal of Environmental Education}, 34\penalty0
  (3):\penalty0 12--21, 2003.

\bibitem[Verplanken et~al.(1998)Verplanken, Aarts, Knippenberg, and
  Moonen]{verplanken1998habit}
B~Verplanken, H~Aarts, A~Knippenberg, and A~Moonen.
\newblock Habit versus planned behaviour: A field experiment.
\newblock \emph{British Journal of Social Psychology}, 37\penalty0
  (1):\penalty0 111--128, 1998.

\bibitem[Vetter(2016)]{vetter2016possibilities}
Max Vetter.
\newblock \emph{Possibilities, boundaries, and consequences of choice
  architecture: The case of green defaults and environmental attitudes}.
\newblock PhD thesis, 2016.

\bibitem[Wassermann and Schneider(2005)]{wassermann2005edibles}
G~Wassermann and F~Schneider.
\newblock Edibles in household waste.
\newblock In \emph{{P}roceedings of the {T}enth {I}nternational {W}aste
  {M}anagement and {L}andfill {S}ymposium, {S}ardinia}, pages 913--914, 2005.

\bibitem[Watson and Meah(2013)]{watson2013food}
M.~Watson and A.~Meah.
\newblock Food, waste and safety: Negotiating conflicting social anxieties into
  the practices of domestic provisioning.
\newblock \emph{{T}he {S}ociological {R}eview}, 60\penalty0 (S2):\penalty0
  102--120, 2013.

\bibitem[Watts and Dodds(2007)]{watts2007}
D.J. Watts and P.S. Dodds.
\newblock Influentials, networks, and public opinion formation.
\newblock \emph{{J}ournal of {C}onsumer {R}esearch}, 34\penalty0 (4):\penalty0
  441--458, 2007.

\bibitem[Weisbuch et~al.(2002)Weisbuch, Deffuant, Amblard, and
  Nadal]{weisbuch2002meet}
G~Weisbuch, G~Deffuant, F~Amblard, and J.P. Nadal.
\newblock Meet, discuss, and segregate!
\newblock \emph{Complexity}, 7\penalty0 (3):\penalty0 55--63, 2002.

\bibitem[Wenlock et~al.(1980)Wenlock, Buss, Derry, and
  Dixon]{wenlock1980household}
R.W. Wenlock, D.H. Buss, B.J. Derry, and E.J. Dixon.
\newblock Household food wastage in {B}ritain.
\newblock \emph{{B}ritish {J}ournal of {N}utrition}, 43\penalty0 (01):\penalty0
  53--70, 1980.

\bibitem[Wood and Hayes(2012)]{wood2012influence}
W~Wood and T~Hayes.
\newblock Social influence on consumer decisions: {M}otives, modes, and
  consequences.
\newblock \emph{{J}ournal of {C}onsumer {P}sychology}, 22\penalty0
  (3):\penalty0 324--328, 2012.

\bibitem[{WRAP}(2014)]{Wrap2014a}
{WRAP}.
\newblock Household food and drink waste: A people focus.
\newblock Banbury, 2014.

\bibitem[Young et~al.(2017{\natexlab{a}})Young, Russell, and
  Barkemeyer]{young2017social}
C.W. Young, S~Russell, and R~Barkemeyer.
\newblock Social media is not the 'silver bullet' to reducing household food
  waste, a response to grainger and stewart (2017).
\newblock \emph{{R}esources, {C}onservation and {R}ecycling},
  2017{\natexlab{a}}.

\bibitem[Young et~al.(2017{\natexlab{b}})Young, Russell, Robinson, and
  Barkemeyer]{young2017can}
W~Young, S.V. Russell, C.A. Robinson, and R~Barkemeyer.
\newblock Can social media be a tool for reducing consumers' food waste? {A}
  behaviour change experiment by a {UK} retailer.
\newblock \emph{{R}esources, {C}onservation and {R}ecycling}, 117:\penalty0
  195--203, 2017{\natexlab{b}}.

\bibitem[Zamri et~al.(2020)Zamri, Azizal, Nakamura, Okada, Nordin, Othman,
  Akhir, Sobian, Kaida, and Hara]{zamri2020delivery}
Gesyeana~Bazlyn Zamri, Nur Khaiyum~Abizal Azizal, Shohei Nakamura, Koji Okada,
  Norul~Hajar Nordin, Nor{\'A}zizi Othman, Fazrena Nadia~MD Akhir, Azrina
  Sobian, Naoko Kaida, and Hirofumi Hara.
\newblock Delivery, impact and approach of household food waste reduction
  campaigns.
\newblock \emph{Journal of Cleaner Production}, 246:\penalty0 118969, 2020.

\bibitem[Zemborain and Johar(2007)]{Zemborain2007}
M.R. Zemborain and G.V. Johar.
\newblock Attitudinal ambivalence and openness to persuasion: {A} framework for
  interpersonal influence.
\newblock \emph{{J}ournal of {C}onsumer {R}esearch}, 33\penalty0 (4):\penalty0
  506--514, 2007.

\end{thebibliography}
\bibliographystyle{plainnat}

\end{document}